\documentclass[aps,prx,twocolumn,superscriptaddress,noeprint,longbibliography, nofootinbib]{revtex4-1}
\usepackage[T1]{fontenc}

\usepackage{graphicx}
\usepackage{bm}
\usepackage{physics}
\usepackage{xcolor}
\usepackage{enumitem}
\usepackage{amsmath, amssymb}
\usepackage[normalem]{ulem}
\usepackage{natbib}
\usepackage{bbm}
\usepackage{comment}

\makeatletter
\let\scshape\relax 
\DeclareRobustCommand\scshape{%
  \not@math@alphabet\scshape\relax
  \ifnum\pdf@strcmp{\f@family}{\familydefault}=\z@
    \fontfamily{paj}%
  \fi
  \fontshape\scdefault\selectfont}
\makeatother

\usepackage[colorlinks=true]{hyperref}
\hypersetup{citecolor = blue}

\newcommand{\unit}{\mathbbm{1}}
\newcommand{\mat}[1]{\textsf{#1}}

\newcommand{\im}{\textrm{i}}

\hypersetup{
    colorlinks=true, 
    linktoc=all,     
    linkcolor=blue,  
    citecolor=magenta
}

\begin{document}

\title{Mesoscopic transport in a Chern mosaic}

\author{Sayak Bhattacharjee}
\email{sayakbhattacharjee@stanford.edu}
\affiliation{Department of Physics, Stanford University, Stanford, CA 94305, USA}
\author{Julian May-Mann}
\affiliation{Department of Physics, Stanford University, Stanford, CA 94305, USA}
\author{Yves H. Kwan}
\affiliation{Department of Physics, University of Texas at Dallas, Richardson, Texas 75080, USA}
\author{Trithep Devakul}
\email{tdevakul@stanford.edu}
\affiliation{Department of Physics, Stanford University, Stanford, CA 94305, USA}
\author{Aaron Sharpe}
\email{aaron.sharpe@stanford.edu}
\affiliation{Department of Physics, Stanford University, Stanford, CA 94305, USA}
\affiliation{Stanford Institute for Materials and Energy Sciences, SLAC National Accelerator Laboratory, Menlo Park, CA 94025}

\begin{abstract}

We analyze mesoscopic electronic transport in a Chern mosaic: a regular pattern of domains whose electronic bands carry differing local Chern numbers. 
An example platform where a Chern mosaic can arise is a moir\'e heterostructure, where variations in the local moir\'e parameters can produce such domains.
We compute resistances at linear response for a variety of domain wall network geometries at zero temperature and magnetic field. 
Simple domain configurations can exhibit zero, integer, or fractional multiples of the quantum of resistance in both the longitudinal and transverse (Hall) responses. 
Our simple semi-classical analysis provides a useful computational method and comparative catalog for ongoing experiments in two-dimensional topological  materials. 
\end{abstract}

\maketitle

\noindent

\section{Introduction}\label{sec: intro}

Two-dimensional, gapped
quantum phases of matter with broken time-reversal symmetry can show a Hall conductance quantized to integer multiples of $e^2/h$ \cite{klitzing1980new, laughlin1981quantized, haldane1988model}. The integer prefactor is given by the total Chern number $(C)$ \cite{thouless1982quantized} of the occupied electronic bands. 
In a macroscopic system characterized by Chern number $C$, such as an integer quantum Hall state, there exist $|C|$ gapless chiral modes at the boundary, which dominate charge transport \cite{klitzing1980new, tsui1982two, chang_experimental_2013, sharpe_emergent_2019, serlin_intrinsic_2020, park_observation_2023, lu_fractional_2024}.   


In addition to macroscopic systems with a fixed Chern
number, it is  possible to realize systems  
that are composed
of multiple mesoscopic scale domains with differing
Chern numbers. This state has been called a \textit{Chern mosaic}~\cite{volovik2018topology, shi_moire_2021, cea_band_2020, mao_quasiperiodicity_2021, devakul2023magic}. Transport in a Chern mosaic is determined by modes not only at the physical boundaries of
the system, but also modes localized at domain
walls in the bulk of the sample, which separate gapped Chern domains~\cite{williams_quantum_2007, abanin2007quantized, amet_selective_2014}. 
This leads to a number
of unusual transport signatures, such as pronounced, but
unquantized, anomalous Hall responses~\cite{chen_tunable_2020, lu_fractional_2024, choi_superconductivity_2025, waters_chern_2025, xie_tunable_2025, chen_electrically_2021, polshyn_electrical_2020, xia_topological_2025, xia_magic_2025, foutty_mapping_2024, li_quantum_2021, park_observation_2023, redekop_direct_2024, xu_observation_2023}. The transport properties of such a 
topological state are the main focus of this study.



In general, a Chern mosaic can arise in any system where there exists a mechanism that can produce spatial variations in the Chern number of the resultant effective local electronic bands. There are a number of possible extrinsic mechanisms that one can consider. Nominally, disorder can induce a Chern mosaic near a percolative transition between two integer quantum Hall states~\cite{chalker1988percolation}. It is also possible to directly imprint domains in certain situations, using electrostatic gating or by intentionally manipulating magnetic domains~\cite{tschirhart2023spinhalltorque,zhang2024manipulation,holtzmann2025opticalcontrolintegerfractional,cai2025opticalswitchingmoirechern,huber2025opticalcontroltopologicalchern}. This has been realized in graphene $p-n$ junctions and magnetic topological insulator devices, respectively~\cite{rosen_chiral_2017, yasuda_quantized_2017}.

A Chern mosaic can also be an intrinsic feature of a system (i.e., not associated with disorder or local manipulation). This is believed to be the case in a number of moir\'e van der Waals (vdW) materials~\cite{sharpe_emergent_2019, serlin_intrinsic_2020, chen_tunable_2020,wang_unusual_2023, bi_designing_2019, de_jong_imaging_2022, nakatsuji_moire_2022}. A typical example is a system that features multiple incommensurate moir\'e patterns, i.e., a multi-moir\'e system~\cite{zhu_twisted_2020, yang_multi-moire_2024}.
Here, spatial variation in the moir\'e pattern over a longer `supermoir\'e' period can produce changes in the local electronic properties, leading to the formation of Chern domains~\cite{zhu_twisted_2020, zhang_spin-polarized_2022, popov_magic_2023, nakatsuji_multiscale_2023,yang_multi-moire_2024, foo_extended_2024, hesp_cryogenic_2024, hoke_imaging_2024, xia_topological_2025, xie_strong_2025, devakul2023magic, kwan_strong-coupling_2024, hoke_linking_2025, xia_magic_2025}. A notable realization of this is magic-angle twisted bilayer graphene aligned to hexagonal boron nitride (MATBG/hBN)~\cite{tschirhart_imaging_2021, grover_chern_2022}. In this case, the supermoir\'e variation is associated with incommensurability between the graphene on graphene moir\'e pattern and the graphene on hBN moir\'e pattern. The existence of a Chern mosaic has also been proposed to explain the transport properties of twisted trilayer graphene (TTG) with particular choices of interlayer angles, which leads to incommensurability between the moir\'e formed by the first and second layers and that of the second and third layers~\cite{xia_topological_2025, xia_magic_2025, devakul2023magic}. For a generic multi-moir\'e system, the precise structure of a Chern mosaic is expected to be highly susceptible to strain and may deviate from  regularity~\cite{tschirhart_imaging_2021, pierce_unconventional_2021, grover_chern_2022, hesp_cryogenic_2024, lai_moire_2025, ji2024local}. Finally, the entropy of the edge modes themselves may result in thermally driven domain formation~\cite{shavit_domain_2022}.
Transport within such samples now depends sensitively on the network of domains within the sample as well as the details of equilibration between edge modes.

Here, we provide a general 
framework to calculate the expected transport characteristics in the semi-classical limit for Chern mosaics. 
We analytically compute the zero-temperature direct-current (dc) resistances within linear response at zero magnetic field. As we detail ahead, Chern mosaics can yield a myriad of transport measurements. These transport measurements are influenced by mesoscopic properties: (a) the domain wall network geometry, (b) the lead locations, (c) the length over which co-propagating modes equilibrate, and (d) any directional bias in the scattering at the gapless junctions where multiple domain walls meet. The last two of these properties are primarily influenced by microscopic details of the electronic band structure of the material. We focus on a simple limit where any directional scattering bias at a junction is absent, and the mode equilibration length is much smaller than the dimensions of a domain (see also Appendices~\ref{mode_scattering} and \ref{validity}).

Unlike the response of a material with a single quantum Hall domain, the response of a Chern mosaic can be called quantized only in a restricted sense. The response we compute is robust to perturbations within the domains of the mosaic (small compared to the energy gap of the Chern domain), or perturbations that do not influence scattering between the modes at the gapless domain walls and scattering junctions. This leads to a variety of unquantized and atypical dc transport characteristics that we detail below.  

We report theoretical predictions for the transport measurements for a wide variety of domain-wall network geometries, including stripes, square, and triangular mosaics. 
We find that Chern mosaics can exhibit distinctive (and perhaps surprising) transport signatures. A few examples include concurrent vanishing longitudinal and Hall resistances, fractional Hall resistance (unlike fractional Hall conductivity, as in the fractional quantum Hall liquid\footnote{We do not claim that fractional Hall conductances cannot be seen in Chern mosaics (for more discussion, see Sec.~\ref{sec: discussion}).}), and higher integer longitudinal resistances (an integer greater than $h/e^2$) even when the Chern mosaic consists of domains with local Chern number $|C|=1$. We find that these responses depend both on the Chern numbers of the domains and the number of domains within the sample. In fact, we find that Chern mosaics can exhibit resistances that discretely interpolate between well-known transport standards (i.e.~from zero to $C^{-1}(h/e^2)$). Our study also highlights a crucial sensitivity to the location of leads within the sample and the parity of the total number of domains along the length and width of the sample. 

The organization of the paper is as follows. In Sec.~\ref{sec: chern_mosaic}, we introduce the Chern mosaic model, characterizing in detail its properties and equivalence classes. In Sec.~\ref{sec: pheno_network model}, we present a Landauer-B\"uttiker formalism that enables us to compute the dc resistances in a Hall bar sample of a Chern mosaic. We present our results, cataloging several Chern mosaic geometries, in Sec.~\ref{sec: results}. A straightforward generalization of our framework can be made to mosaics with counter-propagating (helical) edge modes, which we discuss briefly in Sec.~\ref{sec: helical_edge_modes}. We conclude by discussing our findings in Sec.~\ref{sec: discussion}, giving an outlook towards recent and ongoing experiments in 2D materials.

\tableofcontents

\section{Chern mosaic model}\label{sec: chern_mosaic}

\subsection{Model definition}
\begin{figure}
    \centering
    \includegraphics[width=\columnwidth]{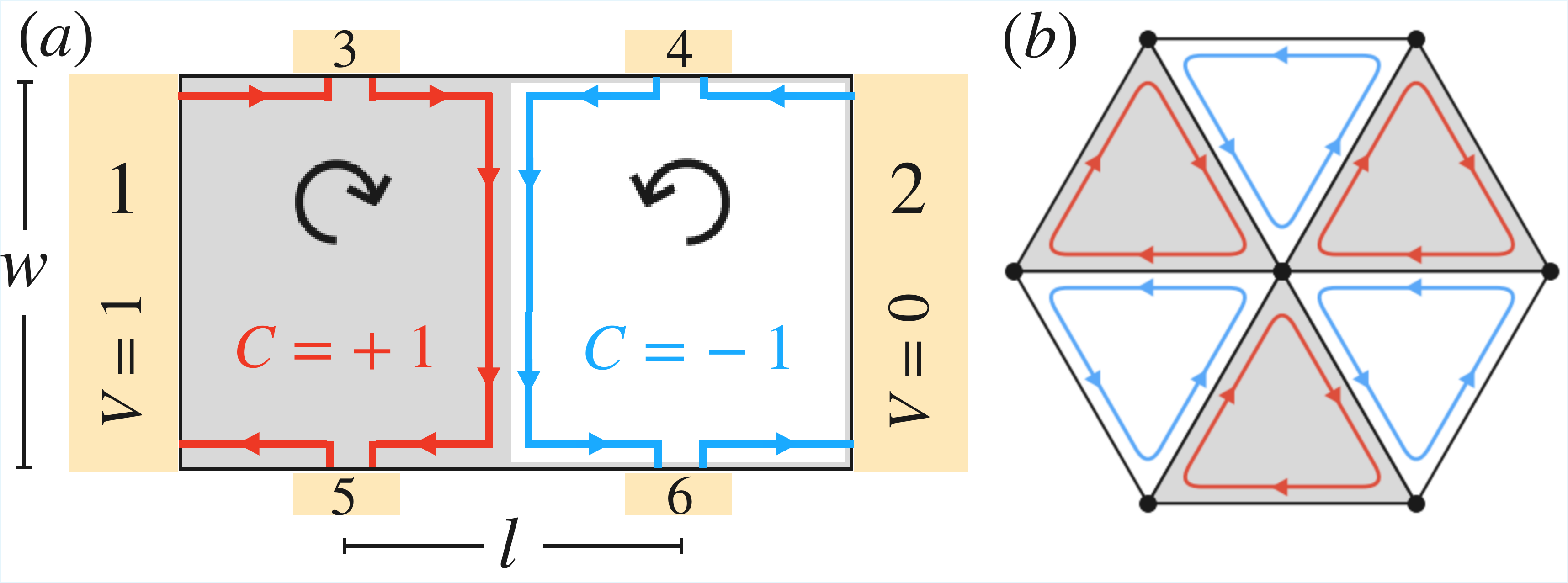}
    \caption{ \textbf{(a)} A schematic Hall bar of a sample with a simple Chern mosaic. The sample has a single vertical domain wall in the center of the Hall bar. The gray domain (left) has Chern number $C=+1$ with a clockwise chiral edge mode (red), while the white domain (right) has a Chern number $C=-1$ with an anti-clockwise chiral edge mode (blue). The leads (light orange) are numbered, and a voltage bias is applied to the source lead ($V=1$) while the sink lead is at voltage $V=0$. Terminals $3-6$ are probe leads. $l$ is the center-to-center distance between the longitudinal probe leads, while $w$ is the width of the Hall bar.   
    \textbf{(b)} Schematic of six $|C|=1$ domains of a triangular Chern mosaic. While the edge modes are shown to not touch at any of the vertices of the triangular domains, physically, they scatter into each other at the scattering junction (black dots at vertices).}
    \label{fig:schematic}
\end{figure}

We define a Chern mosaic as a planar graph embedded on a two-dimensional manifold, 
whose plaquettes have occupied electronic bands with a total non-trivial (local) Chern number. Each plaquette can therefore be regarded as an integer quantum anomalous Hall domain. The edges of the graph are the domain walls, and the vertices are the scattering junctions formed at intersections of domain walls. Open boundary conditions are chosen to be in line with experimental devices, although periodic boundary conditions on an orientable manifold could also be considered. We specifically consider a rectangular Hall bar device geometry, typical for transport measurements (see Fig.~\ref{fig:schematic}(a) for a schematic). Motivated by experiments discussed in Section~\ref{sec: intro}, we focus on \textit{bipartite Chern mosaics} whose plaquettes allow a black-and-white chessboard coloring, i.e.~the dual graph is bipartite.  To set a convention, we assign the black plaquettes Chern number $C_1$ and the white plaquettes Chern number $C_2$, where $C_1, C_2\neq 0$. The bulks of the plaquettes are gapped and are separated by gapless domain walls along which chiral edge modes are present; this allows for a well-defined notion of a Chern number associated to each plaquette. Throughout our work we will not consider trivial insulating plaquettes, although it is possible to extend our analysis to such cases. We assume that the transport properties of the Chern mosaic device are completely determined by a network model of the edge modes and can be analyzed within the Landauer-B\"uttiker framework. The transport coefficients are determined by the properties of the network, subject to boundary conditions of the Hall device (such as orientation of the mosaic with respect to the rectangular boundaries, contact placement, and contact widths). 

First, we characterize equivalence classes of Chern mosaics that exhibit the same transport properties. 
The Chern mosaics in such devices can be extremely sensitive to strain.
For example, small amounts of uniaxial heterostrain in helical trilayer graphene (HTG)---the special case of TTG where the layers are sequentially twisted in the same direction by the same angle---can result in a divergence in one of the moir\'e wavevectors~\cite{hoke_imaging_2024}. Similarly, MATBG/hBN samples realize a somewhat irregular Chern mosaic~\cite{tschirhart_imaging_2021, grover_chern_2022}.
For this study, we therefore consider a wide variety of mosaics: stripes (horizontal and vertical), square lattices and triangular lattices. 
The Chern mosaics of the experiments also have inhomogeneously sized plaquettes, which we ignore---note that the precise geometry of the plaquettes does not influence the transport signatures as long as the network connectivity is preserved. Thus, in the limit of completely equilibrated edge modes and equal mode scattering at scattering junctions, transport is only determined by the connectivity of the graph defined by the domains and we therefore fix the domains to all have the same shape for convenience.

Second, we define the geometrical dimensions of the Hall bar. We consider a Hall bar of linear dimensions $l\times w$. The Hall bar has six contacts---a source lead, a sink lead, and four probe leads (Fig.~\ref{fig:schematic}(a)). The width of the probe leads are denoted by $w_{\textrm{lead}}$. 
We set the source and sink lead width equal to the Hall bar width $w$. We also require the probe lead widths to be less than the domain sizes and, for most of our discussion, we restrict the probe leads to be at most a domain away from the source or the sink (we relax this constraint in Sec.~\ref{subsec: lead_pos}). Since these constraints may not be representative of all experiments, we also provide a voltage map of the network of domain walls of the Chern mosaic so that the resistance across any two domains can be computed. Note that the transport measurements are independent of the precise position of a probe lead along a given domain edge. We will also discuss the case when the probe leads are in contact with multiple domains. We always consider setups where the probe leads on the top boundary and bottom boundary of the Hall bar are aligned, to avoid geometric $R_{xx}-R_{xy}$ mixing in the transport measurements. More specifically, the leads only need to be aligned up to the uncertainty along the $x$-axis that the lead stays in contact with the same domain, since this does not change the transport measurement.

Third, we quantify the number of domains within the Hall bar. Generally, we restrict to mosaics where the axes of the lattices are aligned with the axes of the Hall bar. For square and triangular lattices, we affix a vertex at the bottom-left corner of the Hall bar. This is straightforward for the square Chern mosaic. For the triangular Chern mosaic, our convention is to align any one of the three axes of the triangular lattice (at $\pi/3$ radians with respect to each other) along the $x-$axis of the Hall bar. For stripes, the number of domains is controlled by a single parameter $n$, which counts the number of stripes. For the square and triangular Chern mosaics, the tuplet $(m,n)$ denotes the number of domains along the $x-$axis and the $y-$axis, respectively. 


\subsection{Number of chiral modes on a domain wall}

By the bulk-boundary correspondence for Chern insulators~\cite{hatsugai1993chern, mong2011edge}, the net chirality of edge modes at a domain wall between two plaquettes is given by the difference in Chern numbers of the plaquettes. The total number of edge modes on a domain wall thus depends on both $C_1$ and $C_2$. There are two cases to discuss: (a)  $C_1C_2>0$, where each domain wall has a set of counter-propagating modes, and (b)  $C_1C_2<0$, where all of the edge modes co-propagate on the domain walls. We call such Chern mosaics \textit{unipolar} and \textit{bipolar} respectively, borrowing terminology from quantum Hall devices \cite{abanin2007quantized}.  

In both cases, we assert that the effective number of edge modes supported on the domain wall---which, in fact, co-propagate---is given by $\Delta C = |C_1-C_2|$. In the unipolar case, this occurs because interactions between the edge modes can gap out the counter-propagating modes. The counter-propagating modes should also be spatially proximate to enable efficient inter-mode scattering. Symmetries may prevent interactions from gapping out the edge modes. One example is if the counter-propagating modes carry distinct spin or valley quantum numbers, and the interactions preserve the corresponding flavor symmetries, leading to gapless helical edge channels.
Note that in the unipolar case with $C_1=C_2$, the two domains are equivalent to a single Chern domain with $C=C_1=C_2$ in the regime that the counter-propagating edge modes gap each other out. In the bipolar Chern mosaic, the total number of modes is clearly $|C_1|+|C_2|=|C_1-C_2|$. Particular edges of a plaquette at the boundary of the Hall bar sample have an interface with the vacuum, so the number of edge modes is simply given by the Chern number of the plaquette $|C_1|$ or $|C_2|$. 




In the main text of the paper, we restrict to the simplest bipolar Chern mosaic with integer Chern numbers $C_1=-C_2=1$. Thus, there are two co-propagating edge modes between any two plaquettes in the bulk of the Hall bar, and one on the edge of the sample. In Appendix \ref{app: higher Chern mosaics}, we discuss generalizations to bipartite Chern mosaics with $\Delta C>2$, which we call \textit{higher-order Chern mosaics}.

\subsection{Separation of length scales}

Given the multiple length scales in the Chern mosaic, it is important to make explicit a separation of length scales in the problem. 

The largest length scales of interest are the Hall bar dimensions $l\times w$. Though variable, vdW devices are often designed with $w\sim1\ \mu$m and the separation between contacts is approximately $\gtrsim1$\,$\mu$m. The linear domain size is denoted by $d$. We imagine keeping the total number of domains at most $O(100)$, which should represent experimental devices well~\cite{tschirhart_imaging_2021, pierce_unconventional_2021, grover_chern_2022, hesp_cryogenic_2024, hoke_imaging_2024}.  
We therefore take $d\simeq l/m = w/n$ for the square/triangular lattices, and $d\simeq l/n$ or $d\simeq w/n$ for the stripes depending on their orientation, where $m,n$ are positive integers that are, at most $O(10)$. As we will see in Section~\ref{sec: results}, because we consider taking all modes along a given domain wall to be well equilibrated, the transport properties we calculate only depend upon the number of domains and where the voltage probes are positioned. 

We note that even for a finite $l\times w$, the system is in the thermodynamic limit with respect to the intrinsic lengthscale characterizing the bulk of each domain. For the example of a supermoir\'e Chern mosaic, the relevant lengthscale is the moir\'e period $a_\text{moir\'e}$, which should be much smaller than $d$.
For chiefly academic purposes, we will sometimes also consider the asymptotic limit of our results by taking $n$ (or $m,n$) $\rightarrow \infty$, but this limit is to be interpreted as that obtained by taking $l,w\rightarrow \infty$, so as to keep $d$ fixed.

In moir\'e platforms, a moir\'e pattern modulates interlayer tunneling on a length scale $(a_{\textrm{moir\'e}})$ that is much larger than the atomic length scale $a$.
For MATBG, $\theta\sim1.1^\circ$~\cite{bistritzer2011moire} which corresponds to $a_{\textrm{moir\'e}}\sim13\ \mathrm{nm}$.  
Systems with two super-imposed moir\'es can form a supermoir\'e~\cite{zhu_twisted_2020, zhang_spin-polarized_2022, popov_magic_2023, nakatsuji_multiscale_2023, yang_multi-moire_2024, foo_extended_2024, hesp_cryogenic_2024, hoke_imaging_2024, xia_topological_2025, xie_strong_2025, devakul2023magic, kwan_strong-coupling_2024, hoke_linking_2025, xia_magic_2025}. 
When the layers are twisted in the same direction by the same amount, $a_{\textrm{supermoir\'e}}\simeq a/(2\sin(\theta/2))^2$.
For the special case of TTG where the layers are twisted in the same direction by $1.8^\circ$ (magic-angle helical trilayer [HTG])~\cite{devakul2023magic, popov_magic_2023, nakatsuji_multiscale_2023, yang_multi-moire_2024, xia_topological_2025, xie_strong_2025, devakul2023magic, kwan_strong-coupling_2024,bocarsly2026electrically}, $a_{\textrm{supermoir\'e}}\sim250\ \mathrm{nm}$. For HTG, $d\simeq a_{\textrm{supermoir\'e}}$.
For MATBG/hBN, the Chern number depends on the interlayer shift between the graphene-graphene and graphene-hBN moir\'es~\cite{shi_moire_2021,lin2020doublemoire,cea_band_2020,kwan2021competition}. In principle, these two moir\'es can be exactly commensurate, but twist angle variations in real samples have thus far limited $d\simeq O(10)a_{\textrm{moir\'e}}$. In the discussion ahead, we set $d=1$ and treat the number of domains along an axis as an input parameter into the theory, so that the theory applies universally to an arbitrary Chern mosaic.  

The final length scale of importance is the mode equilibration length. This length depends on the domain wall width $\xi$ 
(see Appendix \ref{mode_scattering}), and for the validity of the network model, we require $d\gg \xi$ such that all low energy degrees of freedom are localized close to the domain walls.  As stated in the previous section, we also set the probe lead width $w_\textrm{lead}\leq d$---in fact, typically we assume the probe leads are point contacts such that $ \xi \ll w_\textrm{lead}\ll d$ in favor of universal results. We note, however, that in the experiments,  often $w_\textrm{lead}\simeq d$---such a possibility is also accounted for in our framework.

\section{Landauer-B\"uttiker formalism}\label{sec: pheno_network model}

The key object of interest for computing transport coefficients is the conductance matrix $\mat{G}$ that relates voltages to currents in a multi-lead device. 
The central relationship within linear response is given by the Landauer-B\"uttiker formula \cite{landauer1957spatial, buttiker1990quantized, datta1997electronic},
\begin{equation}\label{gv=i}
    \mat{G}\mathbf{V}=\mathbf{I},
\end{equation}
where $\mathbf{V}$ is a vector whose entries are the voltages at the leads and $\mathbf{I}$ is the vector whose entries are the currents at the leads. For a Hall bar device with 6 leads, $\mat{G}$ is a $6\times 6$ matrix, and $\mathbf{V}$ and $\mathbf{I}$ are $6\times 1$ vectors. It is useful to number the six leads as shown in Fig. \ref{fig:schematic}A and partition the conductance matrix into sub-blocks as follows,
\begin{equation}\label{conductance_matrix_defn}
    \mat{G}=\left(\begin{array}{cc}
        \mat{G}_{\textrm{s}} & \mat{G}_{\textrm{sp}}\\
        \mat{G}_{\textrm{ps}} & \mat{G}_{\textrm{p}}
    \end{array}\right),
\end{equation}
where we have a $2\times 2$ source-sink block $\mat{G}_\textrm{s}$ and a $4\times 4$ probe lead block $\mat{G}_{\textrm{p}}$. We apply an external voltage difference $V$ across the source and sink leads: $V_1=V$ and $V_2=0$. We also assume that the current through the sample, denoted $I$, is sent through the source and drained through the sink; the probe leads have no net current entering or exiting them. This implies that the voltage vector and current vectors are explicitly given by $\mathbf{V}= (V,\: 0,\: V_3,\: V_4,\: V_5,\: V_6)^\textrm{T}$ and $\mathbf{I}=(I,\: -I,\: 0,\: 0,\: 0,\: 0)^\textrm{T}$, respectively. We set $V=1$ for simplicity throughout the remainder of this work.  

Two properties of the conductance matrix are useful to point out. First, the conductance matrix is a zero-sum matrix due to charge conservation. This implies that the matrix elements of each row and column sum to zero. This sum rule helps constrain the scattering matrix elements---for a lead with $k$ leads from which current flows to the lead in question, specifying the transmission probabilities from $k-1$ leads is sufficient. This sum rule also implies that $\mat{G}$ is a singular matrix, and cannot be na\"ively inverted to compute the voltages. A work-around is to take a pseudo-inverse, for which the resulting non-uniqueness in the voltages is due to the fact that only relative voltages are physical~\cite{bhattacharjee2023green}. Second, applying a time-reversal operation on the Chern mosaic maps the conductance matrix to its transpose (this fact is evident from the Landauer-B\"uttiker formula (Eq. \ref{landauer-buttiker})) and is due to Onsager's reciprocity relation \cite{onsager1931reciprocal, buttiker1988symmetry}. Thus, setting a convention for the coloring of the bipartite Chern mosaic is sufficient for computing $\mat{G}$. 

The transport coefficients that we compute in this study are the longitudinal and transverse (Hall) resistance, denoted by $R_{xx}$ and $R_{xy}$, respectively. We assume we are at zero temperature and zero external magnetic field. These resistances can be computed from the voltage and current vector by the following definition,
\begin{equation}\label{resistance_def1}
    R_{ij}= I^{-1}(V_i-V_j),
\end{equation}
where $i,j$ label the probe leads.  For isotropic systems, $R_{xx}=R_{34}=R_{56}$ and $R_{xy}=R_{35}=R_{46}$, in a Hall bar with leads labeled as in Fig.~\ref{fig:schematic}A.   

For a time-reversal invariant isotropic conductor, $R_{xx}$ is finite but scales with the aspect ratio of the sample while $R_{xy}=0$. One eliminates the dependence on system dimensions by computing the longitudinal and Hall resistivities given by,
\begin{equation}\label{resistance_def2}
    \rho_{xx}=R_{xx}(w/l)\;\;\; \textrm{and} \;\;\; \rho_{xy}=R_{xy}.
\end{equation}
Rotational invariance ensures that these are the only two independent resistivities---$\rho_{xx}=\rho_{yy}$ and $\rho_{xy}=-\rho_{yx}$. 
We note that in this context $l$ is actually the distance between the probe leads, and not the actual length of the Hall bar\footnote{To be more specific, for leads with finite widths, the probe typically averages the potential across its length; in this case, $l$ denotes the center-to-center distance between the probe leads.} (see Fig.~\ref{fig:schematic}(a)). For an integer quantum (anomalous) Hall insulator, $R_{xx}=\rho_{xx}=0$ and $R_{xy}=\rho_{xy}=(h/e^2){C}^{-1}$~\cite{thouless1982quantized}. Throughout the rest of this work, we measure resistances in units of the quantum of resistance $h/e^2$. 

For a finite-sized Chern mosaic, we will find that generically $R_{xx}$ and $R_{xy}$ can both be non-zero and quantized. In such a case, we point out that resistivity is no longer an intrinsic quantity of the system because the quantity $R_{xx}(w/l)$ is \textit{not} independent of $l$ and $w$. Thus, we will present our results in terms of the resistances and not the resistivities. A mosaic-specific definition of resistivity can be defined by factoring out the dependence on the number of domains along the $x$ direction in the longitudinal resistance (see for example Sec.~\ref{sec: n_vertical_stripes} or the discussion in Sec.~\ref{sec: discussion}).

Furthermore, inhomogeneity at the mesoscopic scale implies that $R_{xx}$ is not necessarily the same when measured at the top and bottom boundaries of a Chern mosaic; similarly, $R_{xy}$ is not necessarily the same when measured across different Hall pairs of the sample. This is unlike the typical situation encountered in a homogeneous system, e.g., a normal conductor, trivial insulator, or a quantum Hall insulator. To capture this, for each Chern mosaic Hall bar, we compute each of the following four resistances defined below,
\begin{align}
\begin{split}\label{resitance_def2}
&R_{xx}^{\textrm{top}}=R_{34}, \: R_{xx}^{\textrm{bot}}= R_{56},\\
    &R_{xy}^{\textrm{left}}= R_{35}, \: R_{xy}^{\textrm{right}}= R_{46},
\end{split}
\end{align}
where the superscript denotes whether the longitudinal resistance is measured at the top or bottom of the Hall bar, and whether the Hall resistance is measured at the left or right of the Hall bar.

Determining the above resistances requires the conductance matrix $\mat{G}$.
This may be computed within the Landauer-B\"uttiker formalism \cite{landauer1957spatial, buttiker1990quantized, datta1997electronic} using the formula,
\begin{equation}\label{landauer-buttiker}
    \mat{G}_{ij}=\sum_{p\in i, q\in j}|S_{pq}|^2,
\end{equation}
where $S_{pq}$ is an entry of the scattering matrix such that $|S_{pq}|^2$ is the scattering probability from mode $p$ to mode $q$. 
This formula implies that the $ij^{\textrm{th}}$ entry of the conductance matrix is proportional to the probability of a current transmission from lead $j$ to lead $i$. While this is straightforward to apply for a quantum Hall sample, computing this probability becomes complicated when the current path from lead $i$ to lead $j$ passes through multiple scattering junctions. The actual probability is a function of the number of scattering junctions encountered in all possible current paths between the two leads, the specifics of the transmission rules at the scattering junctions, and the specifics of mode equilibration.  

\begin{figure}
    \centering
    \includegraphics[width=\columnwidth]{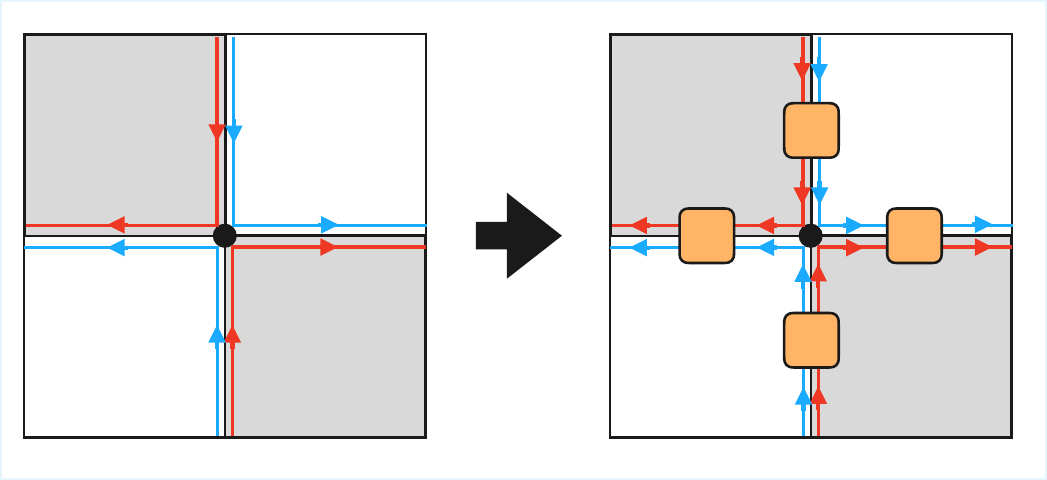}
    \caption{A schematic illustration of adding auxiliary leads (orange rectangles) to the center of the domain walls in (two unit cells of) a square bipolar Chern mosaic with $|C|=1$. The black circle denotes the scattering junction.}
    \label{fig:auxiliary leads added}
\end{figure}

To circumvent this difficulty, we use a convenient formalism that makes the computation straightforward. We introduce auxiliary (fictitious) leads in the bulk of the sample, so that any current sourced at any lead can only scatter once at a single scattering junction before encountering another lead.
A convenient choice of this construction for the Chern mosaic is to place an auxiliary lead on each domain wall (see Fig.~\ref{fig:auxiliary leads added}). 
Upon adding these auxiliary leads to the mosaic, we obtain a generalized form of Eq.~\ref{gv=i} given by, 
\begin{equation}\label{geff_v_i}
\mat{G}_{\textrm{eff}}\mathbf{V}_{\textrm{eff}}=\mathbf{I}_{\textrm{eff}}. 
\end{equation}
Let $N_{\textrm{aux}}$ denote the number of auxiliary leads. Then, $\mat{G}_{\textrm{eff}}$ is an $(N_{\textrm{aux}}+6)\times (N_{\textrm{aux}}+6)$ size matrix and has the following block form,
\begin{equation}\label{g_eff}
\mat{G}_{\textrm{eff}}=\begin{pmatrix}
        \mat{G}_{\textrm{phys}} & \mat{G}_{\textrm{phys,aux}}\\
        \mat{G}_{\textrm{aux,phys}} & \mat{G}_{\textrm{aux}}
    \end{pmatrix},
\end{equation}
where $\mat{G}_{\textrm{phys}}$ is a $6\times 6$ matrix in the basis of the \textit{physical} leads of the Hall bar, while $\mat{G}_{\textrm{aux}}$ is an $N_{\textrm{aux}}\times N_{\textrm{aux}}$ matrix in the basis of the auxiliary leads. The voltage and current vectors are appropriately enlarged and given by $\mathbf{V}_{\textrm{eff}}= (\mathbf{V},\: \mathbf{V}_{\textrm{aux}})^\textrm{T}$ and  $\mathbf{I}_{\textrm{eff}}=(\mathbf{I},\: \mathbf{0})^\textrm{T}$ (we note that the auxiliary leads, like the probe leads, also have no net current entering or exiting). An important point is that $\mat{G}_{\textrm{phys}}\neq \mat{G}$ even though they share the same lead basis. This is because the entries are altered due to the auxiliary leads. Using Eqs.~\ref{geff_v_i} and \ref{g_eff}, and comparing with Eq.~\ref{gv=i}, we obtain the following  algebraic formula for $\mat{G}$,
\begin{equation}\label{g_formula}    \mat{G}=\mat{G}_{\textrm{phys}}-\mat{G}_{\textrm{phys,aux}}(\mat{G}_{\textrm{aux}})^{-1}\mat{G}_{\textrm{aux,phys}}.
\end{equation}
This is a key formula of our work, and implies that given a rule to compute $\mat{G}_{\textrm{eff}}$, the resistances can be straightforwardly computed using Eqs.~\ref{g_formula} and \ref{resistance_def1}.  This formula can be interpreted as follows:  $\mat{G}_{\textrm{phys}}$ is the contribution to the conductance matrix from modes that directly couple one lead to another, while $\mat{G}_{\textrm{phys,aux}}(\mat{G}_{\textrm{aux}})^{-1}\mat{G}_{\textrm{aux,phys}}$ captures processes that involve one or more scattering events. 

The remaining task is to determine the entries of $\mat{G}_{\textrm{eff}}$. Generically, they depend on the details of mode scattering and equilibration. However, as motivated earlier, we consider the limit where (a) the modes have completely equilibrated by the time they reach a scattering junction, and (b) when scattering at the junction, they have equal probability to scatter into any of the available outgoing edge modes. In this limit, using the Landauer-B\"uttiker formula (Eq.~\ref{landauer-buttiker}), we obtain the rule for the entries of the conductance matrix. A phenomenological treatment of relaxing these assumptions is provided in Appendix~\ref{mode_scattering}. The zero-sum nature of the matrix constrains the entries, which helps determine the scattering rules for leads at the edge of the sample. We discuss explicitly the rules for the square and triangular Chern mosaics, while the rule for the striped Chern mosaic can be easily inferred from this discussion. First, consider a lead $i$ in the bulk of the sample which also exclusively receives current from leads only in the bulk of the sample. The corresponding matrix element in the conductance matrix is given by,
\begin{equation}\label{formula_eff_G} [\mat{G}_{\textrm{eff}}]_{ij}= \begin{cases}
        & n_{\textrm{out}}\;\;\; \textrm{if} \; i=j\\
        & -n_{\textrm{out}}/\alpha \;\;\; \textrm{if}\; i\neq j \:\textrm{and}\:  \textrm{path}_{j\rightarrow i}=1\\
        & 0 \;\;\; \textrm{otherwise}.
    \end{cases},
\end{equation}
Here 
$n_{\textrm{out}}$ is the number of modes coming out of the lead $i$, $\textrm{path}_{j\rightarrow i}=1$ denotes that there exists a path from lead $j$ to lead $i$ either directly or via a single scattering junction. $\alpha$ is the number of leads (labeled by $j$) that lead $i$ receives current from that are related to $i$ such that $\textrm{path}_{j\rightarrow i}=1$. 
For the square lattice Chern mosaic with $C_1=-C_2=1$, $n_{\textrm{out}}=2$ and  $\alpha=2$, while for the corresponding triangular lattice mosaic $n_{\textrm{out}}=2$ and  $\alpha=3$.  For the square mosaic, this rule applies to all leads except for those on the edge of the sample, in which case the diagonal matrix element is $1$ and so the non-zero off-diagonal matrix element (corresponding to the lead that sources this lead) is $-1$. 

\begin{figure}
    \centering
    \includegraphics[width=0.9\columnwidth]{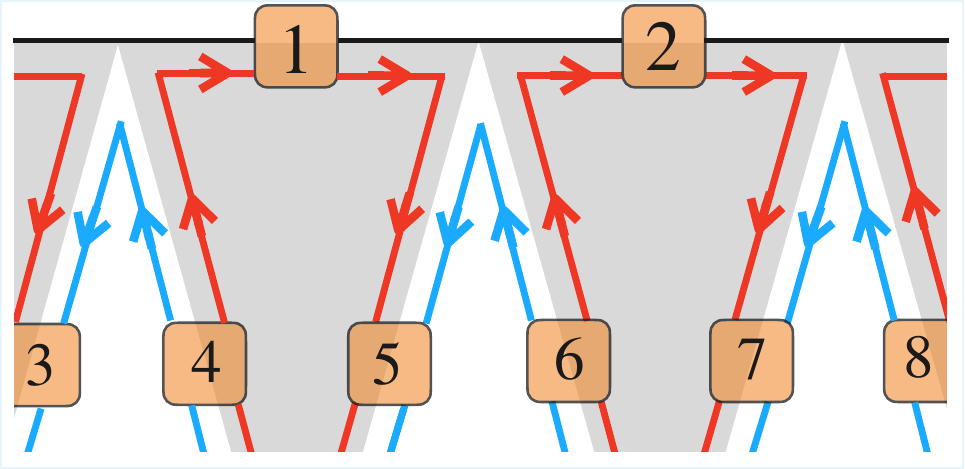}
    \caption{A schematic illustration of a part of a horizontal edge of a Hall bar with a triangular bipolar Chern mosaic with $|C|=1$. The auxiliary leads (dark orange) are numbered 1 through 8. The black horizontal line indicates the horizontal edge of the sample. While the edge modes do not touch at the scattering junctions at the sample edge in the schematic, physically, they scatter into each other.}
    \label{fig:edgetriangular}
\end{figure}

For a triangular mosaic, we wish to explicitly specify the matrix elements of the conductance matrix corresponding to two kinds of leads that are near the edge of the sample. First, consider a lead on the edge of the sample (see e.g. leads 1 and 2 in Fig.~\ref{fig:edgetriangular} that are sourced by both a lead in the row of leads immediately adjacent to the edge \textit{and} a lead on the edge to its left or right). In this case, we assign the diagonal matrix element of such a lead to be 1, while the off-diagonal matrix element corresponding to the lead on the edge is $-1/3$, and the off-diagonal matrix element corresponding to the lead in the row adjacent to the edge is $-2/3$. This assignment of off-diagonal entries is in proportion to the number of modes supplied by these leads, which are 1 and 2 for the lead at the edge and in the row adjacent to the edge respectively. Next, consider an auxiliary lead in the mosaic that is in a row immediately adjacent to the physical edge of the sample. Some of those leads (see e.g. leads 3, 5 and 7 in Fig.~\ref{fig:edgetriangular}) are sourced both by a lead in the same row \textit{and} a lead on the physical edge. However, the lead in the same row supplies twice as many modes as the lead on the edge. Therefore, while the diagonal matrix element for such a lead is $2$, the off-diagonal matrix elements are $-4/3$ and $-2/3$ corresponding to the lead in the same row and the lead on the edge, respectively. One may check that this overall assignment of the edge and auxiliary leads respects the sum-rule discussed previously. Finally, we note that we have ignored scattering processes at a junction touching the source/sink---the presence of the contact shorts the voltage across the scattering junction. This is a heuristic assignment of matrix elements; in a physical sample, these matrix elements will depend on microscopic parameters.

The above construction yields a zero-sum matrix for all the mosaics being considered. The diagonal entries are interpreted as the probability for a mode to scatter to the lead it comes from, while the off-diagonal entries are the probabilities of transmission between distinct leads. For any other mosaics, this construction enables a straightforward extension.  

\section{Results}\label{sec: results}
\begin{figure*}
    \centering
\includegraphics[width=\textwidth]{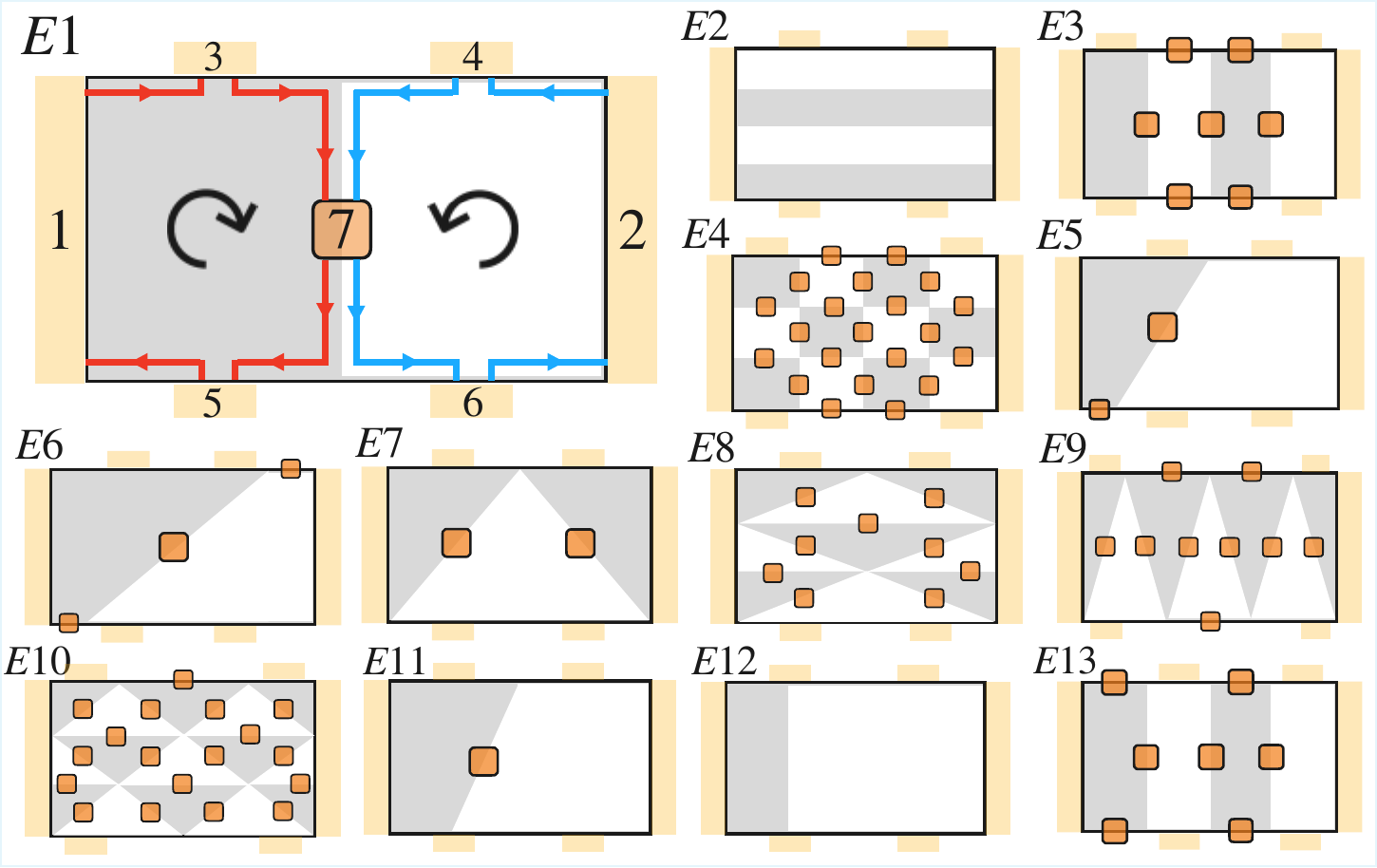}
    \caption{Schematic Hall bars of samples with mosaic geometries whose transport measurements are analytically computed in Sec. \ref{subsec: analytic_examples}. The physical leads are depicted in light orange, while the auxiliary leads on the domain walls are shown in dark orange. We depict representative examples of the mosaics while the text discusses results for general number of domains: e.g., E2 is the $n=4$ horizontally striped mosaic, E4 is the $m=n=4$ square Chern mosaic, E8 is the $n=3$ horizontally striped triangular mosaic (a column of $2\times 3$ triangular domains), E9 is the $n=3$ vertically striped triangular mosaic (a row of $2\times 3$ triangular domains) and E10 is the $m=2,n=3$ triangular mosaic. The black plaquettes carry a clockwise chiral mode while the white plaquettes carry an anti-clockwise chiral mode.}
    \label{fig:examples}
\end{figure*}

In this section, we discuss the results obtained from our model of Chern mosaics. We consider various simple geometries such as striped, square, and triangular Chern mosaics and compute the resistances analytically. 
We remind the reader that we restrict to the bipolar Chern mosaic with $C=\pm 1$ throughout this section. 
The bipartite nature of the Chern mosaic implies that the resistances are sensitive to odd-even effects in the number of the domains, and we thus explicitly discuss both cases whenever relevant.
Our convention in all of the geometries considered is that the bottom-left domain is colored \textit{black}, and thus has $C=+1$; we assume that this domain has edge modes traveling clockwise around the plaquette. The numbering of the source/sink and probe leads follows that shown in Fig. \ref{fig:examples}. For the mosaics with triangular domains, we found it convenient to often state the number of \textit{full} triangular domains---a full triangular domain is a triangular domain with three acute angles. In this language, an erstwhile full triangular domain cut in half by a Hall bar boundary has to be joined with its complement elsewhere in the mosaic to be counted as a full triangular domain. Thus, E7 in Fig. \ref{fig:examples} has 2 full triangular domains, E8 has 6 full triangular domains and so on.   

While we discuss each example in detail below, the reader may refer to Table~\ref{main table} for a catalog of our results.
The corresponding Chern mosaics have been schematically depicted in Fig.~\ref{fig:examples}. The resistances for some of the triangular Chern mosaics do not have simple fractional expressions in terms of the number of domains, and so have been shown in Figs.~\ref{fig:row_resistances} and \ref{fig:c6_resistances}. For these cases, we obtain asymptotic analytical expressions of the resistances in the limit of large number of columns of domains, and we further demonstrate a rapid convergence to these analytical expressions in Fig.~\ref{fig:row_resistances} and Appendix~\ref{small n deviations}. We also depict a complete voltage map of the square and triangular Chern mosaic domain wall networks in Fig.~\ref{fig:voltage_map}. Table~\ref{main table} and Figs.~\ref{fig:row_resistances}, \ref{fig:c6_resistances} and \ref{fig:voltage_map} constitute the main results of our work.

\begin{ruledtabular}
\begin{table*}
 \centering 
\begin{tabular}{lllllll}
Geometry& Voltage $\mathbf{V}^{\textrm{T}}$ & Current $I$ &  $R_{xx}^{\textrm{top}}$ &  $R_{xx}^{\textrm{bot}}$ & $R_{xy}^{\textrm{left}}$ & $R_{xy}^{\textrm{right}}$ \\
\hline

E1. One vertical domain wall & $\left(1\: 0\: 1\: 0\: \frac{1}{2}\: \frac{1}{2}\right)$  & $\frac{1}{2}$ & 2 & 0 & 1& $-1$ \\

E2. $n$ horizontal stripes ($n$ odd) & $\left(1\: 0\: 1\: 1\: 0\: 0\right)$  & $n$ & 0 & 0 & $\frac{1}{n}$& $\frac{1}{n}$ \\
E2. $n$ horizontal stripes ($n$ even) & $\left(1\: 0\: 0\: 0\:0\:0\right)$  & $n$ & 0 & 0 & 0& 0 \\
E3. $n$ vertical stripes ($n$ odd) & $\left(1\: 0\: 1\: \frac{1}{n}\:\frac{n-1}{n}\:0\right)$  & $\frac{1}{n}$ & $n-1$ & $n-1$ & $1$& $1$\\
E3. $n$ vertical stripes ($n$ even) & $\left(1\: 0\: 1\: 0\:\frac{n-1}{n}\:\frac{1}{n}\right)$  & $\frac{1}{n}$ & $n$ & $n-2$ & $1$& $-1$\\
\hline
E4. $(m,n)$ square Chern mosaic ($m$ odd, $n$ odd) & $\left(1\: 0\: 1\: \frac{1}{m}\:\frac{m-1}{m}\:0\right)$  & $\frac{n}{m}$ & $\frac{m-1}{n}$ & $\frac{m-1}{n}$ & $\frac{1}{n}$& $\frac{1}{n}$\\
E4. $(m,n)$ square Chern mosaic ($m$ odd, $n$ even) & $\left(1\: 0\: \frac{m-1}{m}\: 0\:\frac{m-1}{m}\:0\right)$  & $\frac{n}{m}$ & $\frac{m-1}{n}$ & $\frac{m-1}{n}$ & $0$& $0$\\
E4. $(m,n)$ square Chern mosaic ($m$ even, $n$ odd) & $\left(1\: 0\: 1\: 0\:\frac{m-1}{m}\:\frac{1}{m}\right)$  & $\frac{n}{m}$ & $\frac{m}{n}$ & $\frac{m-2}{n}$ & $\frac{1}{n}$& $-\frac{1}{n}$\\
E4. $(m,n)$ square Chern mosaic ($m$ even, $n$ even) & $\left(1\: 0\: \frac{m-1}{m}\: \frac{1}{m}\:\frac{m-1}{m}\:\frac{1}{m}\right)$  & $\frac{n}{m}$ & $\frac{m-2}{n}$ & $\frac{m-2}{n}$ & $0$& $0$\\
E4. One vertical and horizontal domain wall & $\left(1\: 0\: \frac{1}{2}\: \frac{1}{2}\: \frac{1}{2}\: \frac{1}{2}\right)$  & $1$ & 0 & 0 & 0& 0 \\
\hline
E5. One tilted domain across one lead & $\left(1\: 0\: 1\: 0\:\frac{1}{2}\:\frac{1}{2}\right)$  & $\frac{1}{2}$ & $2$ & $0$ & $1$& $-1$\\
E6. One tilted domain across both leads &  $\left(1\: 0\: 1\: 1\: \frac{1}{2}\: \frac{1}{2}\right)$  & $\frac{1}{2}$ & 0 & 0 & 1& $1$\\
\hline
E7. Two tilted domain walls &  $\left(1\: 0\: 1\: \frac{1}{3}\: 1\: 1\right)$  & $\frac{4}{3}$ & $\frac{1}{2}$ & 0 & 0& $-\frac{1}{2}$\\
E8. One column of $2n$ triangular domains ($n$ odd) &  $\left(1\: 0\: 1\: \frac{1}{3}\: 1\: 1\right)$  & $\frac{4n}{3}$ & $\frac{1}{2n}$ & $0$ & $0$& $-\frac{1}{2n}$\\
E8. One column of $2n$ triangular domains ($n$ even)&  $\left(1\: 0\: 1\: 1\: 1\: 1\right)$  & $\frac{4n}{3}$ & 0 & 0 & $0$& $0$\\
E9. One row of $2n$ triangular domains (at large $n$)& $(1\: 0\: 1 \: \frac{3}{6n+1} \: 1\: \frac{6}{6n+1})$ &  $\frac{9}{6n+1}$ & $ \frac{6n-2}{9}$& $\frac{6n-5}{9}$ & 0 & $-\frac{1}{3}$  \\
\hline
E10. $(m,n)$ triangular Chern mosaic (at large $m$ and odd $n$) & $(1\: 0\: 1 \: \frac{3}{6m+1} \: 1\: \frac{6}{6m+1})$  & $\frac{9n}{6m+1}$ & $ \frac{6m-2}{9n}$& $\frac{6m-5}{9n}$ & 0 & $-\frac{1}{3n}$\\
E10. $(m,n)$ triangular Chern mosaic (at large $m$ and even $n$) & $(1\: 0\: 1 \: \frac{6}{6m+1} \: 1\: \frac{6}{6m+1})$  & $\frac{9n}{6m+1}$ & $ \frac{6m-5}{9n}$& $\frac{6m-5}{9n}$ & 0 & $0$\\
\hline
E11. One tilted domain wall in contact with lead 5 &  $\left(1\: 0\: \frac{1}{2}\: 0\: \frac{1}{2}\: \frac{1}{2}\right)$  & $\frac{1}{2}$ & 1 & 0 & 0& $-1$\\
E12. One vertical domain wall in contact with leads 3 \& 5 &  $\left(1\: 0\: \frac{1}{2}\: 0\: \frac{1}{2}\: \frac{1}{2}\right)$  & $\frac{1}{2}$ & 1 & 0 & 0& $-1$\\
\end{tabular}
\caption{Voltages, currents and resistances of the Hall bar for various mosaic geometries considered in \ref{subsec: analytic_examples} and depicted in Fig.~\ref{fig:examples}. The Chern mosaics are bipartite and bipolar and have $|C|=1$, and the transport is computed at zero field and zero temperature  under assumptions of equal mode scattering at the junctions and complete mode equilibration at the domain walls. We fix the source and sink voltages to $1$ and $0$ respectively, which uniquely fixes the current $I$ by Eqs.~\ref{gv=i}, \ref{geff_v_i} and \ref{g_formula}.}
\label{main table}
\end{table*}
\end{ruledtabular}



\subsection*{Illustrative examples}\label{subsec: analytic_examples}


\subsubsection{One vertical domain wall (E1)}\label{subsec: vertical stripes_n=2}

The first example we consider is the Chern mosaic with one vertical domain wall between the two pairs of probe leads. In this case, there is a single mode propagating with opposite handedness in the domains, resulting in two co-propagating modes on the domain wall (see Fig.~\ref{fig:examples} E1). 

We add an auxiliary lead at the center of the domain wall to account for the mode scattering at the two junctions and obtain the following effective conductance matrix (of size $7\times 7$), 
\begin{equation}\label{effective_conductance_matrix}
    \mat{G}_{\textrm{eff}}= \left(\begin{array}{cccccc|c}
        1& & & & -1& &\\
        &1 & & & & -1&\\
        -1& &1 & & & &\\
        & -1& &1 & & &\\
        & & & &1 & &-1\\
        & & & & &1 &-1\\
        \hline 
        & &-1 &-1 & & &2
    \end{array}\right),
\end{equation}
where the blank entries are 0 and the blocks denoted by the solid lines follow those in Eq.~\ref{g_eff}.   Using Eq.~\ref{g_formula}, we compute the conductance matrix of the system, given by,
\begin{equation}\label{reduced_G}
    \mat{G}=\left(\begin{array}{cc|cccc}
          1& & & & -1&\\
        &1 & & & & -1\\
        \hline 
        -1& &1 & & &\\
        & -1& &1 & &\\
        & &-1/2 &-1/2 &1 & \\
        & &-1/2 &-1/2 & &1 
    \end{array}\right),
\end{equation}
where the blocks follow those defined in Eq.~\ref{conductance_matrix_defn}. From this, we readily obtain the resistances using Eqs.~\ref{resistance_def1} and \ref{resitance_def2}, and find that,
\begin{align}
\begin{split}
&\mathbf{V}=\left(1\: 0\: 1\: 0\: \frac{1}{2}\: \frac{1}{2}\right)^\textrm{T},\;\; I=1/2,\\
&R_{xx}^{\textrm{top}}=2,\: R_{xx}^{\textrm{bot}}=0,\: R_{xy}^{\textrm{left}}=1 \;\; \textrm{and}\;\; R_{xy}^{\textrm{right}}=-1.
\end{split}
\end{align}
Already from this simple example, we note how the behavior of the Chern mosaic is unlike that of a conductor, trivial insulator, or quantum Hall insulator, because we see an integer Hall response along with finite longitudinal resistances (albeit only for $R^{\textrm{top}}_{xx}$ in this case). We also note the lack of rotational invariance is manifest in the resistance measurement. 

\subsubsection{$n$ horizontal stripes (E2)}\label{n vertical stripes}
The next geometry we discuss is that of $n$ horizontal stripes. This geometry does not require any auxiliary leads due to the lack of scattering junctions (see Fig.~\ref{fig:examples} E2). This example however provides the opportunity to examine odd-even effects in the number of domains, which we demonstrate below. 

The simplicity of the geometry allows us to compute resistances even without explicitly determining the conductance matrix. Because the leads on the top and bottom are each on the same domain edge, $R_{xx}=0$. Furthermore, the propagation of the edge modes implies that for odd $n$, $V_3=V_4=1$ and $V_5=V_6=0$ while for $n$ even, $V_3=V_4=V_5=V_6=0$. To compute the current, note that for all $n$, there are $n$ modes propagating between the source and sink, which implies $I=n$. This implies that for odd $n$, we obtain a \textit{fractional} Hall resistance of $1/n$ (see also Table \ref{main table}). For even $n$, all resistances are zero. 

We note the distinct behavior of samples with odd and even number of stripes in that for an even number of horizontal stripes, the sample mimics the transport signatures of a superconductor, while for an odd number of stripes, one obtains the Hall response of a Chern insulator with a higher Chern number $|C|>1$.



\subsubsection{$n$ vertical stripes (E3)}\label{sec: n_vertical_stripes}
We next consider a geometry with $n$ vertical stripes (see Fig.~\ref{fig:examples} E3). Here, we must add $n-1+2(n-2)$ auxiliary leads to account for scattering at the junctions on the edge of the sample. In this case, we find that the Hall resistance is $\pm$ 1, while the longitudinal resistances increase linearly with $n$. 

In the limit that $n\rightarrow \infty$, the longitudinal resistance diverges while the Hall resistance continues to be equal to $1$. Na\"ively, this transport signature may resemble that of a \textit{Hall insulator}~\cite{kivelson1992global} (distinct from a quantum Hall state) where $\rho_{xx}\rightarrow \infty$ but $\rho_{xy}$ is finite. However, we note that the divergence is in the resistance and not in the resistivity (which is not well-defined). In fact, an ad hoc definition of a longitudinal resistivity for this mosaic may be defined as follows: $\rho_{xx}=R_{xx}(w/l)$, where we take $l=n$ and $w=1$ (since there is effectively only one mode traveling from the source to the sink), so that $\rho_{xx}=1$ as $n\rightarrow \infty$. This helps draw the distinction with the dc transport signature of the Hall insulator (of course, their microscopic physics is different and would manifest in other physical observables). One can also define a resistivity for a finite sized system that specifically accounts for the mosaic structure. In this case, we may take $l=n$ and then retain a $O(1/n)$ finite-size correction that depends upon whether the resistance is at the top or bottom and whether $n$ is even or odd (see Table \ref{main table}). Due to this mosaic specificity in this definition of the resistivity, we mostly refrain from presenting results in terms of this quantity ahead.  



\subsubsection{$(m,n)$ square Chern mosaic (E4)}\label{mn square}
Now we consider the square Chern mosaic with $m$ columns and $n$ rows (see Fig.~\ref{fig:examples} E4). In this case, we consider the four subcases based on whether $m$ and $n$ are odd or even, and tabulate the resistances in Table~\ref{main table}. 

First, consider the limit of one vertical and one horizontal domain wall $(m=n=2)$. Here, we find zero longitudinal and Hall resistances, akin to the dc transport signature of a superconductor. Having $m,n>2$ guarantees a non-zero longitudinal resistance, which increases linearly with the number of domains between the leads ($m-1$) and decreases inversely with the number of rows in the mosaic $(n)$. Non-zero fractional Hall resistances are found when $n$ is odd---the fraction is set by the number of rows in the mosaic. In the asymptotic limit that $m,n\rightarrow \infty$, the Hall resistance $\rightarrow 0$ while the longitudinal resistance (if finite) approaches the ratio $m/n$. 

For completeness, we briefly consider defining a resistivity (as in the previous subsection) for the square Chern mosaic. Take, for example, the case where $m$ and $n$ are both odd, then $\rho_{xx}=R_{xx}(w/l)$, and we must take $w=n$ (since there are $n$ modes effectively going from source to sink). To scale out the dependence on $m$, we can take $l=m-1$, so that we find a $\rho_{xx}=1$ both at the top and bottom. Such a construction may be carried out for other mosaics as well and typically yields a quantized $\rho_{xx}=1$, but as noted before, the definition of resistivity is mosaic specific. 

\subsubsection{One tilted domain wall (E5, E6)}
In general, domain walls of a Chern mosaic do not need to be parallel to the axes of the Hall bar. We now consider mosaic geometries where the domain wall is tilted with respect to the axes of the Hall bar. For a single domain wall, two topologically distinct scenarios arise---(a) where the domain wall passes between one pair of aligned probe leads (see Fig.~\ref{fig:examples} E5), and (b) where the domain wall passes between both pairs of aligned probe leads (see Fig.~\ref{fig:examples} E6). We choose a geometry where the tilted domain walls do not touch the source/sink. This allows for two scattering events both on the top and bottom edge of the sample. The consequence of the source/sink leads touching a domain wall has been considered before, for example, in the horizontally striped mosaic or the square Chern mosaic. To account for the scattering junctions at the edges of the sample, we insert an auxiliary lead at the domain wall and on any of the remaining domain edges on the edge of the sample. Note that in these geometries, strictly speaking, we break our convention of always computing the resistances of probe leads exactly next to the source/sink. However, this situation is relevant for a Hall bar sample where the (shorter) distance between the scattering junction and the source/sink is 
significant enough that the scattering junction is not shorted.

Having inserted these auxiliary leads, it is obvious that both E5 and E6 are equivalent to E1---it is just that the resistances are measured across different leads. Then, one can read off the voltages from those found for E1, and we tabulate the resistances in Table~\ref{main table} (see Ref.~\cite{finney_extended_2025} for a similar computation). An important takeaway from this exercise is that a finite $R_{xx}$ is associated with at least one domain wall between the two leads across which the longitudinal measurement is made.

\subsubsection{Two tilted domain walls (E7)}
The tilted domain wall geometry considered above naturally extends to the triangular Chern mosaic. As a preliminary example, we first discuss the geometry where the sample contains two full triangular domains---in particular, the two tilted domain walls meet at a scattering junction on the edge, between the two probe leads (see Fig.~\ref{fig:examples} E7). In this case, we allow the tilted domain walls to touch the source and the sink, and as mentioned in the beginning of Sec.~\ref{sec: pheno_network model}, their effect is to short out the scattering junctions there. Note that the geometry related to this by $C_2$ symmetry can be obtained by simply renumbering the probe leads, as will be true for any of the $C_2$ asymmetric mosaics considered in this section. We tabulate the resistances in Table~\ref{main table}---we find fractional $R_{xx}$ and $R_{xy}$ (with magnitude $1/2$). Note that the introduction of the scattering junction at the edge where four domain walls meet (which was absent in the single tilted domain wall examples in E5 and E6) causes such a fractional response (unlike the integer resistances in E5 and E6).

\begin{figure}
    \centering
\includegraphics[width=\columnwidth]{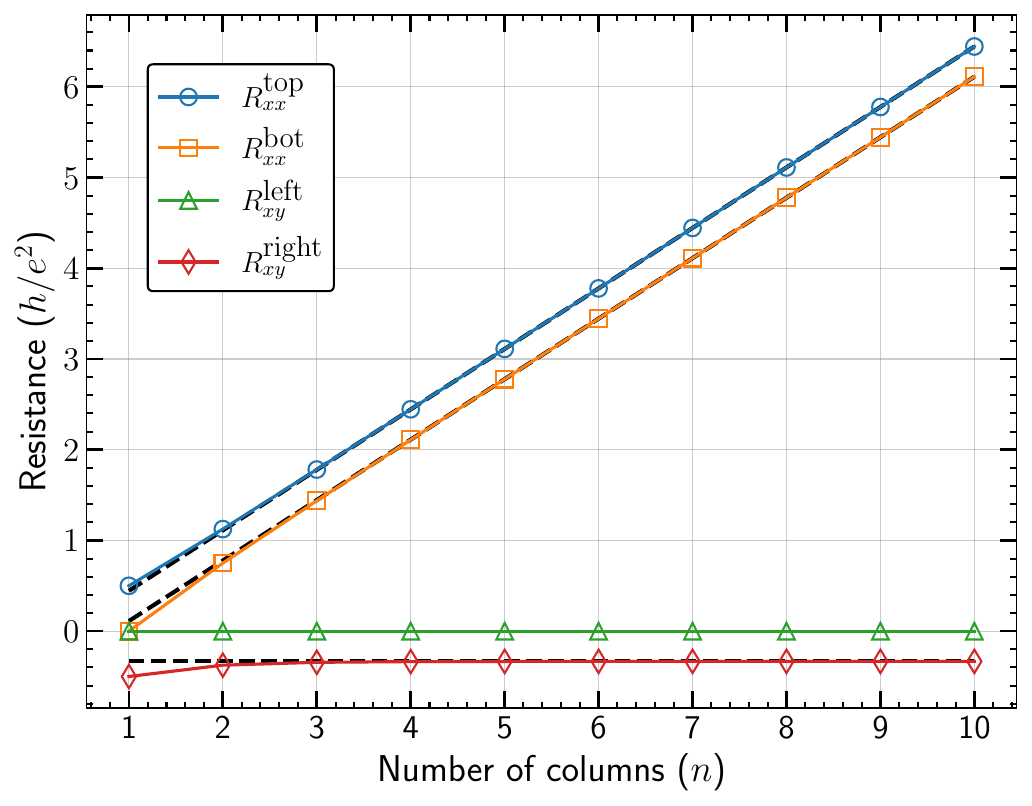}
    \caption{Resistances of the mosaic with a row of triangular domains with $n$ columns ($2n$ full triangular domains)---see Fig.~\ref{fig:examples} E9. Here, $n=1,2,\hdots, 10$. Note that $R_{xy}^{\textrm{left}}=0$ for all $n$. $R_{xx}^{\textrm{top}}$ and $R_{xx}^{\textrm{bot}}$ monotonically increase with $n$, just as in the mosaic  in E3, with $n$ vertical stripes. The black dashed lines represent the asymptotic values of the resistances for large $n$, given in Table~\ref{main table} (E9)---$R_{xx}^{\textrm{top}}=(6n-2)/9$, $R_{xx}^{\textrm{bot}}=(6n-5)/9$ and $R_{xy}^{\textrm{right}}=-1/3$.} 
    \label{fig:row_resistances}
\end{figure}

\subsubsection{One column or row of $2n$ triangular domains (E8, E9)}
First, consider a column of triangular domains in a Chern mosaic as shown in Fig.~\ref{fig:examples} E8. This has $n$ rows, and $2n$ complete triangular domains. We are able to compute the resistances in closed form (as a function of $n$) for this mosaic, and tabulate it in Table~\ref{main table}. Note that when $n$ is even, the resistances are zero while when $n$ is odd, the non-zero longitudinal and Hall resistances are equal in magnitude and given by $R=1/2n$. This matches with our discussion in the previous subsection, which is the limit when $n=1$. 

Next, we consider a row of triangular domains---$n$ columns so that there are $2n$ triangular domains (see Fig.~\ref{fig:examples} E9). Here, the resistances are not simple fractions, although they can be computed analytically. We plot the resistances for $n=1$ to $10$ in Fig.~\ref{fig:row_resistances}. We find that $R_{xy}^{\textrm{left}}=0$, while $R_{xx}$ increases monotonically with $n$. Note that again, the $n=1$ limit matches with the resistances found in the previous subsection. 

\begin{figure*}
\centering
    \includegraphics[width=1\textwidth]{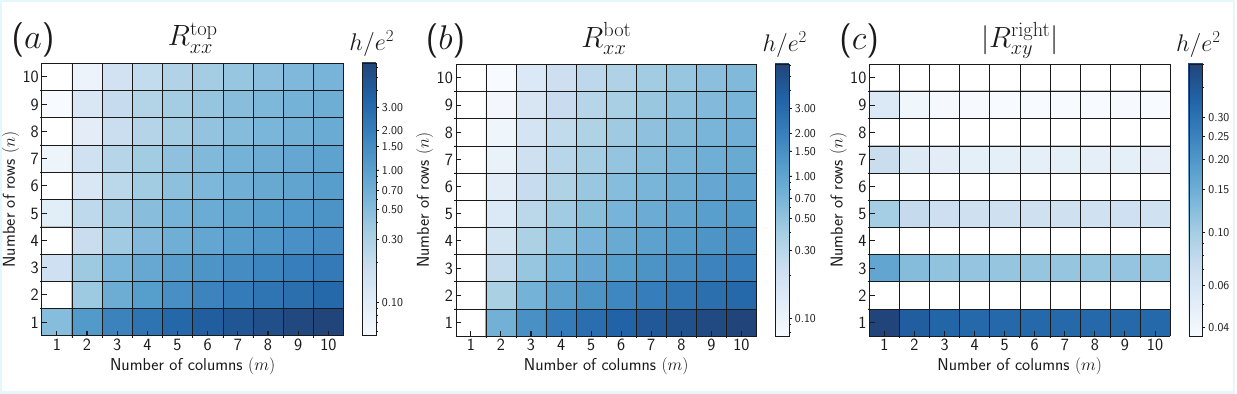}
    \caption{Resistances of the triangular Chern mosaic as a function of the number of columns, $m$,  and the number of rows, $n$ (see Fig. \ref{fig:examples} E10). There are $2mn$ triangular domains in the Hall bar. \textbf{(a), (b)} and \textbf{(c)} show $R_{xx}^{\textrm{top}}$, $R_{xx}^{\textrm{bot}}$ and $-R_{xy}^{\textrm{right}}$, respectively, while $R_{xy}^{\textrm{left}}=0$ for all $(m,n)$. We show resistances for $m,n=1,2,\hdots, 10$. Note that the colorbars display resistance magnitudes in units of $h/e^2$ on a logarithmic scale. The trends are summarized in Sec. \ref{triangular_chern}. Asymptotic expressions for large $m$ are tabulated in Table~\ref{main table}.}
    \label{fig:c6_resistances}
\end{figure*}

We find, however, a rapid convergence of the non-zero resistances to the following asymptotic resistance values: $R_{xx}^{\textrm{top}}=(6n-2)/9$, $R_{xx}^{\textrm{bot}}=(6n-5)/9$ and $R_{xy}^{\textrm{right}}=-1/3$. To compare the actual resistances to these asymptotic expressions, we have also plotted these asymptotics with dashed lines in Fig.~\ref{fig:row_resistances}. The closed form of the corresponding voltages and current is also provided in Table~\ref{main table} (E9). In Appendix~\ref{small n deviations}, we provide the explicit values of the voltages and further quantify the deviations at small $n$ from these asymptotic values. For $n=4$, the deviation is already only $\approx 0.02\%$.

\subsubsection{$(m,n)$ triangular Chern mosaic (E10)}\label{triangular_chern}

We now consider an $(m,n)$ triangular Chern mosaic (see Fig.~\ref{fig:examples} E10). This implies that we consider $m$ columns and $n$ rows in the Chern mosaic, so that there are a total of $2mn$ triangular domains. Like for the row of triangular domains, it is not possible to find simple fractions of $m$ and $n$ for the resistances, so we instead plot the resistance values in Fig.~\ref{fig:c6_resistances}. The asymptotic forms for large $m$ are, however, obtainable and tabulated in Table~\ref{main table} (E10). We discuss this more below. We find that $R_{xy}^{\textrm{left}}=0$ for all $(m,n)$. The longitudinal resistances increase with increasing number of columns and decrease with increasing number of rows while exhibiting odd-even effects. The Hall resistance on the right is non-zero only when $n$ is even, and decreases with increasing number of columns. As $m,n\rightarrow \infty$, from the asymptotic expressions, we find that the triangular Chern mosaic is metallic (like the square counterpart), with a longitudinal resistance of $2m/3n$. 

A simple limit that is of interest is when there is a single $C_6$ symmetric scattering junction in the bulk of the sample---this occurs when $m=1$ and $n=2$. This sample has four complete triangular domains. Since the top and bottom probes lie in the same domain, the longitudinal resistance is 0, and we find the Hall resistance is 0 as well (see Fig.~\ref{fig:c6_resistances}), similar to the mosaic with one vertical and one horizontal domain wall discussed in Sec.~\ref{mn square}. 

The asymptotic expressions of the large $m$ asymptotes are similar to those for the $(m,n)$ square Chern mosaic in the following sense: the voltages are independent of the number of rows $(n)$ and thus directly obtainable from the result of a single row with $n$ columns. Given that this is true for both the square and triangular mosaics, we believe that this may be a generic property of high-symmetry Chern mosaics (such as lattices) and may be derivable from Eq.~\ref{g_formula}. The effect of increasing the number of columns is to increase the total current proportionally (by a factor of $(n)$). We note that a deviation from a simple fraction at small $n$ is not seen in the square Chern mosaic. Thus, we suspect that the deviation in the triangular mosaic may be due to boundary effects that may be attributed to the non-alignment of the rectangular geometry of the Hall bar with the natural axes of the triangular mosaic. Note that no deviation is found in the current of the $(m,n)$ triangular mosaic for small $n$ (number of rows). This is consistent with our guess, because the longer edge of the Hall bar is aligned with the axis along the rows of the triangular Chern mosaic. In Appendix~\ref{oblique_hall_bar}, we consider an oblique Hall bar with boundaries aligned with the axes of a triangular Chern mosaic, and obtain a resistance measurement expressible as a simple fraction; this supports the above conjecture.

\subsubsection{Domain wall in contact with a lead (E11, E12)}\label{domain wall_contact_lead}
In general, a probe lead can be in contact with a domain wall. Such a situation is particularly likely to arise for Chern mosaics in moir\'e platforms where the probe lead width can be comparable to a domain size. 
Our framework also allows us to compute transport in such a configuration.
We have shown earlier how Eq.~\ref{formula_eff_G} applies to situations when a lead touches a domain wall---for instance, in the horizontally striped Chern mosaic, a horizontal domain wall is in contact with the source and sink. As simple examples, we return to the case of a single domain wall and consider two scenarios---(a) a tilted domain wall in contact with a single lead (specifically lead 5), with the other end of the domain wall in between the two probe leads on the other edge (see Fig.~\ref{fig:examples} E11), and (b) a vertical domain wall in contact with a pair of probe leads (specifically leads 3 and 5) (see Fig.~\ref{fig:examples} E12). 

Only the tilted domain wall needs an auxiliary lead, while the conductance matrix for the vertical domain wall can be directly written due to the lack of scattering junctions. We tabulate these  resistances in Table~\ref{main table}. In both cases, we find the same voltage distribution at the leads, yielding the same resistances. Comparing E11 and E12 with their closest mosaic counterparts where a probe is not in contact with a domain wall, namely E5 and E1, respectively, we see a similar change in the measured resistances. $R_{xx}^{\textrm{top}}$ has decreased from 2 to 1 and $R_{xy}^{\textrm{left}}$ has been shorted out due to the probe contacting the domain wall. 

\subsubsection{Probe leads away from source/sink (E13)}\label{subsec: lead_pos}

\begin{figure}
    \centering
    \includegraphics[width=\columnwidth]{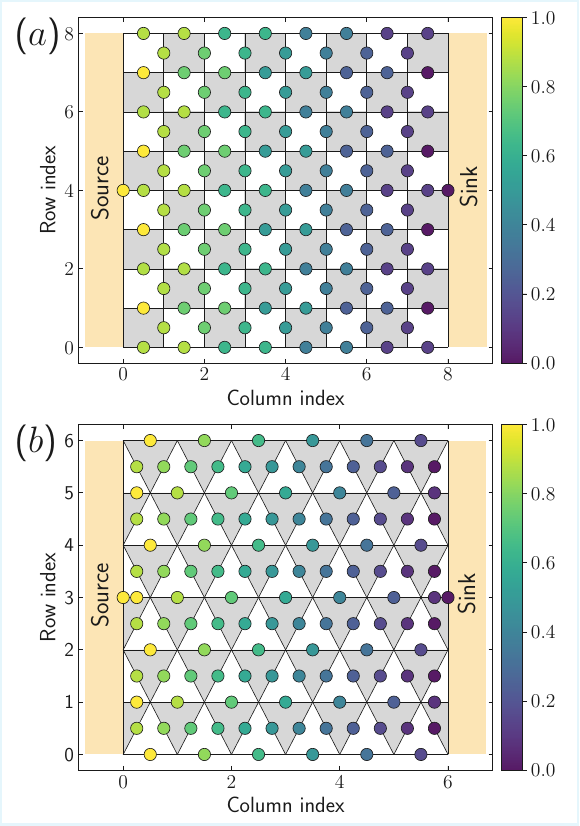}
    \caption{Voltage map for the \textbf{(a)} $(8,8)$ square and \textbf{(b)} $(6,6)$ triangular Chern mosaic with $|C|=1$. The gray (white) domains supply a clockwise (anti-clockwise) edge mode to the domain walls. The voltages are depicted on the gapless domain wall network formed due to the Chern mosaic by colored blobs. The voltage at the source $V=1$, while at the sink $V=0$. We find that the current $I=1$ for the square mosaic, while for the triangular mosaic $I=1.459469$ (rounded off to six decimal places).}
    \label{fig:voltage_map}
\end{figure}

While the discussion so far has considered a wide variety of geometries, we have mostly restricted to placing the probe leads at most a domain away from the source and the sink. This may not be the case in experiments, so we now discuss how the resistances change as a function of distance of the probe leads from the source/sink (or the distance between the longitudinal leads). For example, one may consider a Hall bar of the kind in Fig.~\ref{fig:examples} E13, where there are 4 vertical stripes but the left Hall pair is not immediately next to the source. For this discussion, we find it convenient to simply plot the voltage distribution of the Chern mosaic. We obtain the voltages at the auxiliary leads---this gives the voltage map of the domain wall network. The voltage distribution in the insulating bulks of the domains should be the solution of Laplace's equation subject to the voltage boundary conditions on the domain wall. Thus, we essentially obtain the complete voltage map of the Chern mosaic. In particular, we discuss the square and triangular Chern mosaics. After placing the auxiliary leads, the enlarged conductance matrix $\mat{G}_{\textrm{eff}}$ can be inverted to obtain the voltages throughout the mosaic---namely, from Eq.~\ref{geff_v_i}. The voltage maps are plotted in Fig.~\ref{fig:voltage_map} for an $8\times 8$ square Chern mosaic and a $6\times 6$ triangular Chern mosaic domain wall network. The resistances can now be computed between any pair of probe leads on the edge of the sample---$R=\Delta V/I$. As usual, we set $V=1$ at the source and $V=0$ at the sink. We find that the current $I=1$ for the square mosaic, while it is equal to $I=1.459469$ (rounded off to six decimal places) for the triangular mosaic, which is $\approx 54/37$, in agreement with the closed form expression.

\section{Mosaics with helical edge modes}\label{sec: helical_edge_modes}

\begin{figure}
    \centering
    \includegraphics[width=\columnwidth]{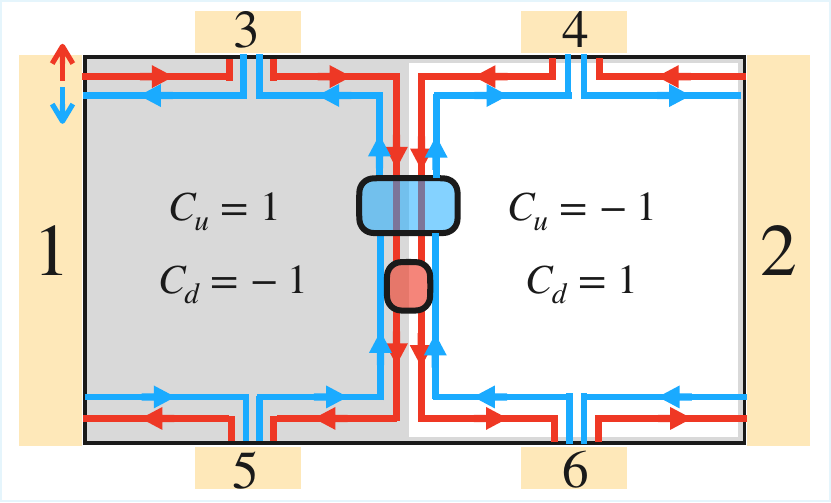}
    \caption{A schematic of a Chern mosaic with a pair of helical edge modes per domain wall in the bulk. Red and blue denote the up and down spin channels respectively. $C_u$ and $C_d$ denote the up and down spin Chern number respectively. Note the need for two auxiliary leads on a domain wall without a physical lead, to account for separate voltages of each spin channel. The red (blue) auxiliary lead interacts only with the up (down) spin. The resistances for this mosaic are provided in Eq.~\ref{resistances_spinhall}.}
    \label{fig:spinhall}
\end{figure}

Our formalism can be straightforwardly generalized to mosaics with counter-propagating (helical) edge modes. This is relevant to experiments; for example, in the absence of
symmetry-breaking, marginally twisted bilayer graphene~\cite{huang2018topologically, rickhaus2018transport}
and helical trilayer graphene~\cite{xia_topological_2025, hoke_imaging_2024} have a pair of helical edge modes per domain wall (per spin) protected
by approximate valley-$U(1)$ symmetry. 

For concreteness, we consider a bipartite mosaic with two pairs of helical edge modes per domain wall in the bulk. The helical edge modes can have opposite spin polarizations, and arise from separate conservation of each spin channel, which enforces counter-propagating modes with opposite spin quantum numbers along the domain walls. That is, we may consider an up spin-Chern number of $+1$ on the black plaquettes and an up spin-Chern number of $-1$ on the white plaquettes, and the opposite for the down spin-Chern number (see Fig.~\ref{fig:spinhall} for an example mosaic). Again, we will restrict our discussion to the limit of equal branching and complete mode equilibration, but within each spin channel. In this case, we find that the conductance matrix $\mat{G}$ of the six terminal Hall bar can be obtained using the following relation,
\begin{equation}\label{qsh_conductance}
\mat{G}=\mat{G}_{\textrm{Chern}}+\mat{G}_{\textrm{Chern}}^\textrm{T},
\end{equation}
where $\mat{G}_\textrm{Chern}$ is the conductance matrix corresponding to the transport of one of the spin channels, say, the up-polarized spin. This is reminiscent of the relationship of the transport of a quantum spin Hall insulator to that of the Chern insulator, suggesting that such a mosaic may be thought of as two time-reversed copies of the appropriate Chern mosaic. 

To derive Eq.~\ref{qsh_conductance}, we construct an effective conductance matrix $\mat{G}$, but now we must add a pair of auxiliary leads on each domain wall without a physical lead in the sample, because the two spin channels can generically have different voltages (see Fig.~\ref{fig:spinhall}). Only at the \textit{physical} leads at the edge of the sample (the contacts) do the two spin channels equilibrate. The rule for obtaining the entries of this conductance matrix is the same as that for the Chern mosaic, with the constraint that the up-spin and down-spin channels do not mix. A convenient arrangement of the effective conductance matrix is into blocks labeled by the spin channels, as follows,
\begin{equation}
\mat{G}_{\textrm{eff}}=\left(\begin{array}{ccc}
\mat{G}_{\textrm{phys}}& \mat{G}_{\textrm{phys,aux}}^u& \mat{G}_{\textrm{phys,aux}}^d\\
\mat{G}_{\textrm{aux,phys}}^u& \mat{G}_{\textrm{aux}}^u& 0\\ 
\mat{G}_{\textrm{aux},\textrm{phys}}^d& 0& \mat{G}_{\textrm{aux}}^d
\end{array}\right),
\end{equation}
where the submatrices have the same meaning as in Eq.~\ref{g_eff}, and the index $u,d$ denotes the spin label (up and down respectively). This matrix can be decomposed into the spin channels as follows,
\begin{align}\begin{split}
\mat{G}_\textrm{eff}&=\mat{G}_{\textrm{eff},u}+\mat{G}_{\textrm{eff},d},\\
\mat{G}_{\textrm{eff},u}&=\left(\begin{array}{ccc}
\mat{G}_{\textrm{phys}}& \mat{G}_{\textrm{phys,aux}}^u& 0\\
\mat{G}_{\textrm{aux,phys}}^u& \mat{G}_{\textrm{aux}}^u& 0\\ 
0& 0&0\\ 
\end{array}\right),\\ \mat{G}_{\textrm{eff},d}&=\left(\begin{array}{ccc}
\mat{G}_{\textrm{phys}}&0 & \mat{G}_{\textrm{phys,aux}}^d\\
0& 0& 0\\ 
\mat{G}_{\textrm{aux},\textrm{phys}}^d& 0& \mat{G}_{\textrm{aux}}^d\\ 
\end{array}\right).
\end{split}
\end{align}
To obtain the $6\times 6$ conductance matrix $\mat{G}$ in Eq.~\ref{qsh_conductance}, we must solve Eq.~\ref{geff_v_i}, where  $\mathbf{V}_{\textrm{eff}}=(\mathbf{V}\: \mathbf{V}_{\textrm{aux},u} \: \mathbf{V}_{\textrm{aux},d} )$ and $\mathbf{I}_{\textrm{eff}}=(\mathbf{I}\: \mathbf{0}\:\mathbf{0})$, and $\mathbf{V}$ and $\mathbf{I}$ are the voltage and current vectors of the physical leads with their usual definitions. Then, we obtain the generalization to Eq.~\ref{g_formula}, given by,
\begin{equation}\label{generalized_g_eff_formula}
\mat{G} = \mat{G}_{\textrm{phys}}-\sum_{\alpha=u,d}\mat{G}_{\textrm{phys,aux}}^\alpha[\mat{G}_{\textrm{aux}}^\alpha]^{-1}\mat{G}_{\textrm{aux,phys}}^\alpha.
\end{equation}
This may be interpreted similar to Eq.~\ref{g_formula}: the conductance matrix for the Hall bar now contains corrections coming from scattering within each spin channel, with the contributions acting independent of each other due to the conservation of each spin channel. Decomposing $\mat{G}_\textrm{phys} \:(=\mat{G}_{\textrm{phys},u}+\mat{G}_{\textrm{phys},d})$ further into the two spin channels, we obtain Eq.~\ref{qsh_conductance} via the following identifications,
\begin{align}\begin{split}&\mat{G}_\textrm{Chern}=\mat{G}_{\textrm{phys},u}-\mat{G}_{\textrm{phys,aux}}^u[\mat{G}_{\textrm{aux}}^u]^{-1}\mat{G}_{\textrm{aux,phys}}^u,\\
&\mat{G}_\textrm{Chern}^\textrm{T}=\mat{G}_{\textrm{phys},d}-\mat{G}_{\textrm{phys,aux}}^d[\mat{G}_{\textrm{aux}}^d]^{-1}\mat{G}_{\textrm{aux,phys}}^d.
\end{split}\end{align}
Note that since the conductance matrix is a symmetric matrix, the Hall resistances of a mosaic with helical edges modes must vanish. As an illustrative example, we compute the longitudinal resistances for the simplest mosaic with helical edge modes, one with a single vertical domain wall carrying two pairs of helical edge modes, using the formalism outlined above. The matrix $\mat{G}$ can be obtained using Eq.~\ref{qsh_conductance}, where $\mat{G}_\textrm{Chern}$ is given in Eq.~\ref{reduced_G}. We obtain, 
\begin{align}
\begin{split}\label{resistances_spinhall}
&\mathbf{V}=\left(1\: 0\: \frac{3}{4}\: \frac{1}{4}\: \frac{3}{4}\: \frac{1}{4}\right)^\textrm{T},\;\; I=\frac{1}{2},\\
&R_{xx}^{\textrm{top}}= R_{xx}^{\textrm{bot}}=1,\: R_{xy}^{\textrm{left}} =R_{xy}^{\textrm{right}}=0.
\end{split}
\end{align}
In this simple case, we find the auxiliary lead voltages (the red and blue lead on the vertical domain wall in Fig.~\ref{fig:spinhall}) are both equal to $1/2$, but this is not generically true for mosaics with helical edge modes. 

\section{Discussion}\label{sec: discussion}

In this paper, we developed a general framework for computing mesoscopic electronic transport in Chern mosaics. 
A primary use-case of our model is to serve as a comparison between the controlled limit of an idealized Chern mosaic, and the transport properties of experimental systems that could host a Chern mosaic. 
Although the parameter space is large---requiring consideration of both the microscopic details of the system in addition to the mosaic structure itself---our framework can be applied to infer possible domain structures that would be consistent with the observed transport. If additional constraints can be applied, such as the Chern numbers of domains through band structure calculations, it may be possible to uniquely determine the domain structure from transport alone.  

Using this framework, we identified specific Chern mosaics that exhibit the following transport signatures: 
\begin{enumerate}
    \item $R_{xx}=R_{xy}=0$, mimicking the zero temperature dc transport of a superconductor. This vanishing Hall resistance may occur when the Chern numbers of the domains cancel out on average, such as when there exists an equal number of  $C=+1$ and $C=-1$ domains (see e.g. E2 ($n$ even), E4 ($m=n=2$) and E8 ($n$ even) in Table \ref{main table}).  Note that this is a necessary but not sufficient condition; consider for example, the vertical striped Chern mosaic with $n$ even (see Table~\ref{main table}).  
    \item $R_{xx}=0, R_{xy}=n^{-1}(h/e^2)$, where $n$ is the number of domains in a mosaic with horizontal striped domains. This is seen in E2 ($n$ odd) in Table \ref{main table}. This is the same as the transport of an integer quantum Hall state with multiple fully filled Landau levels, and yet can be seen in a mosaic with only $|C|=1$ domains. 
    \item $R_{xx}=\gamma(h/e^2), R_{xy}=0$, where $\gamma$ is a rational number dependent on the number of domains along the $x$ and $y$ axes. See e.g. E4 ($m$ odd, $n$ even or $m,n$ both even and $\neq 2$) in Table \ref{main table}. This is also true for the mosaic with helical edge modes discussed in Sec.~\ref{sec: helical_edge_modes}. 
    \item $R_{xx}=\gamma_1(h/e^2), R_{xy}=\gamma_2(h/e^2)$, where $\gamma_1$ and $\gamma_2 \:(<1)$ are  rational numbers dependent on the number of domains along the $x$ and $y$ axes. See e.g. E4 ($m,n$ both odd or $m$ even, $n$ odd) in Table~\ref{main table}. 
\end{enumerate}
While the third signature resembles that of a normal metal, the fourth signature appears to be unlike the transport of conventional phases of matter (normal metals, trivial insulators or quantum Hall states). 
Our four conclusions above are also unaffected in the presence of domains with Chern numbers $|C|>1$ (see Appendix~\ref{app: higher Chern mosaics}). 
We have also pointed out that due to mesoscopic scale inhomogeneity, the usual definition of resistivity does not hold for such systems.
Even if a notion of resistivity can be defined (as noted in Sec.~\ref{sec: n_vertical_stripes}), we find that it is mosaic geometry-specific, and may be useful to only scale out the domain number dependence between the two leads measuring the longitudinal resistance.

Our results highlight the importance of experimental signatures capable of differentiating a Chern mosaic from other phases. 
To distinguish a Chern insulator from a Chern mosaic, the former would exhibit a gap whose density–field slope follows a Str\v{e}da relation $\, \partial n/\partial B=Ce/h$ consistent with the observed value of Hall resistance \cite{streda1982theory}. 
Differentiating a superconductor from a Chern mosaic that exhibits zero longitudinal and Hall resistance is more subtle.
The breakdown of edge modes under applied current can mimic the non-linear transport characteristics typically associated with superconductivity~\cite{komiyama_breakdown_1985, eaves_size-dependent_1986, nachtwei_breakdown_1999, komiyama_heat_2000, berdyugin_out--equilibrium_2022, fox_part-per-million_2018, rosen2022measured}. 
Further supportive evidence of a superconductor would be oscillations of the critical current under an out-of-plane magnetic field characteristic of phase coherent transport. 
While very clean moir\'e systems can exhibit single-particle interference phenomena, such as Fabry-P\'erot and Aharonov–Bohm oscillations~\cite{rickhaus_transport_2018,xu2019giant}, these interference phenomena should persist to much higher fields and should only weakly depend on the applied current in contrast to the critical-current oscillations of an expected superconductor.
Furthermore, a quantized longitudinal resistance in dc transport can indicate either the presence of helical modes or a gapless domain wall between the relevant contact pair. 
Absent definitive signatures of a particular phase, more reliable conclusions require consideration of data from multiple closely spaced Hall probes. 


We identify three natural extensions of our work. (a) A microscopic modeling of the gapless network of the Chern mosaic. In Appendix~\ref{mode_scattering}, we discuss how phenomenologically one can relax equal mode scattering and equilibration. These parameters can be obtained systematically by capturing phase coherence of the edge modes. Such a calculation of the transport of the appropriate (clean) Chalker-Coddington network \cite{chalker1988percolation} may, for example, help analyze Aharonov-Bohm oscillations in relatively clean Chern mosaics (for similar calculations, see Refs.~\cite{xu2019giant, de2020aharonov}). (b) This ties into computing finite field transport. Within our model, this can be implemented in a phenomenological manner by noting that at low-fields, semi-classically, the effect of the Lorentz force is to introduce a directional bias at the scattering junctions. (c) Finally, finite temperature transport---given a knowledge of the conductance matrix $\mat{G}$, this may be computed using the finite temperature Landauer-B\"uttiker formula~\cite{datta1997electronic}. At low temperatures, the conductance matrix we obtain continues to hold. In fact, thermally activated mode equilibration further justifies the  assumptions in our analysis. At higher temperatures, a more involved analysis may be required to study inter-mode scattering and activation of bulk carriers, in which case, our network model ceases to apply.  

We note that our calculation assumes a network model for the Chern mosaic, so that no charge transport occurs through the bulk of the incompressible domains. However, recent experiments in magnetically doped topological insulators~\cite{rosen2022measured,ferguson2023direct} and moir\'e materials~\cite{ji2024local} show that the position of the conduction channels can meander away from the edge, depending on sample microscopics and gate electrostatics. We note that in such a limit, the Landauer-B\"uttiker formula (Eqs.~\ref{g_formula} and \ref{formula_eff_G})---even if it yields the correct answer for a single domain~\cite{douccot2024meandering}---may not work here due to reduced (or lack of) inter-mode equilibration.  

Fractional filling of topological bands in 2D materials has realized fractional quantum (anomalous) Hall states~\cite{neupert2011fci,regnault2011fci,sheng2011fractional,sun2011nearly,tang2011high,spanton2018observation,xie2021fractional,cai2023signatures,zeng2023thermodynamic,park_observation_2023,xu_observation_2023,lu_fractional_2024,xie_tunable_2025,choi_superconductivity_2025,waters_chern_2025,aronson2025displacement,lu2025extended}.
This motivates the notion of a \textit{fractional Chern mosaic}, where each of the domains is a fractional quantum (anomalous) Hall phase~\cite{kwan2024fractional}. Consider a fractional Chern mosaic where the domains are equivalent to the $1/3$ Laughlin states~\cite{laughlin1983anomalous} (with alternating handedness for the chiral edge modes in adjacent domains)---in this case, our analysis easily applies and the resistances are thrice the resistances of the $|C|=1$ bipartite Chern mosaic. This statement assumes that the two chiral edge modes carrying $e/3$ charge each behave similarly to the modes for the integer quantum Hall domains, without edge mode reconstruction~\cite{wan2003edge,chamon1994sharp}. For fractional quantum Hall domains with edge reconstruction or more intricate edge mode structure~\cite{kane1994randomness, levin2007particle}, a more involved analysis is required. Note that while our results so far for the (integer) Chern mosaic show only zero or fractional Hall resistances with a fraction less than or equal to one, such a fractional Chern mosaic can show a Hall resistance with a higher integer or fractional resistances greater than one as well. On accounting for unequal mode scattering, this can be obtained in the integer Chern mosaic as well (see Appendix \ref{mode_scattering} and Ref. \cite{finney_extended_2025}).

Finally, the voltage plots for the Chern mosaic plotted in Fig.~\ref{fig:voltage_map} are reminiscent of the solution to a discretized version of Poisson's equation subject to appropriate boundary conditions. However, this is not the standard Laplace's equation due to the handedness of the current paths of the chiral edge modes. We find it useful to interpret the Chern mosaic network then as a \textit{diode} network; this is also apparent from the Landauer-B\"uttiker relation $\mat{G}_{\textrm{eff}}\mathbf{V}_{\textrm{eff}}=\mathbf{I}_{\textrm{eff}}$, which is simply a lattice discretization of Poisson's equation for diode networks (a similar formulation for a resistor network is in Ref.~\cite{bhattacharjee2023green}, for example). We point out that analyzing the percolation properties of such a disordered \textit{chiral} diode network may be an interesting open question~\cite{redner1982directed}, with applications to transport in dirty Chern mosaics.

\acknowledgements

SB acknowledges discussions with Chaitanya Murthy and Srinivas Raghu. We thank David Goldhaber-Gordon, Jesse Hoke and Sandesh Kalantre for reading the manuscript. SB is supported in part by the US Department of Energy, Office of Basic Energy Sciences, Division
of Materials Sciences and Engineering, under Contract No. DE-AC02-76SF00515S. AS is supported by the US Department of Energy, Office of Science, Basic Energy Sciences, Materials Sciences and Engineering Division, under Contract DE-AC02-76SF00515. JMM is supported by a Leinweber Institute for Theoretical Physics
fellowship. TD acknowledges support from the Sloan Research Fellowship.

\appendix

\section{Mode scattering and equilibration}\label{mode_scattering}
In the main text, we considered the limit of complete mode equilibration along the domain walls and equal mode scattering at the scattering junctions. In this appendix, we discuss a phenomenological model for mode scattering and equilibration and the limit of that model which applies to the above assumption. Our discussion will closely follow that in Ref.~\cite{takei2016spin}. 



Mode scattering and equilibration can occur broadly through three mechanisms: (a) momentum relaxation by impurity scattering in the gapless domain wall regions, (b) a change in other quantum numbers, such as magnetic disorder mixing modes of opposite spin polarizations or intervalley scatterers mixing modes of different valley polarizations and (c) spatial proximity between the modes, which favors transmission and equilibration due to greater overlap of their wavefunctions. The Chern mosaic has two kinds of scattering junctions: those either in the bulk of the sample or at the edge of the sample. This implies that scattering rules at junctions on the edge of the sample may be more susceptible to sample edge roughness and edge boundary conditions (which for graphene devices, for instance, can be armchair or zigzag)\footnote{Note however that the physical scenario in a Hall bar may be more complicated---when gates are laterally aligned with the physical edge of the sample, field focusing at the edge may screen the physical edge, forming an electrostatically defined edge further inside the sample~\cite{silvestrov_charge_2008}.}. Generically, mode equilibration lengths can also be different for a domain wall as compared to an edge of the sample. To simplify the discussion, let us restrict to Chern mosaics with $|C|=1$. Here, inter-mode equilibration only occurs at the domain walls and not at the physical edge of the sample. The extent of overlap can play an important role in determining transmission probabilities for mode scattering---for instance, in a triangular Chern mosaic, a scattering to a domain wall at $\pi/3$ radians can have different probability to a scattering to a domain wall at $\pi$ radians \cite{efimkin2018helical}. For mode equilibration, we thus require that $\xi \lessapprox  $ twice the width of the edge mode wavefunction, to allow for large overlap of edge mode wavefunctions.  


For mode equilibration, we label the two modes on the domain wall by $1$ and $2$. We assume a constant tunneling conductance per unit length $g$ between the two modes and note that $\delta i_{1,2}(x)=\mp g[i_1(x)-i_2(x)]\delta x$ due to charge conservation (where $x$ is the distance traveled along the domain wall, and $i$ is the current of the mode). This yields a pair of differential equations for the modes,
\begin{equation}
    \frac{\partial i_1(x)}{\partial x}=-    \frac{\partial i_2(x)}{\partial x}=-g[i_1(x)-i_2(x)],
\end{equation}
which admits the solution,
\begin{equation}\label{mode_equilibration_formula}
    \begin{pmatrix}
        i_1(x)\\
        i_2(x)
    \end{pmatrix}=\frac{1}{2}\begin{pmatrix}
        1+e^{-x/\lambda}& 1-e^{-x/\lambda} \\
        1-e^{-x/\lambda}&  1+e^{-x/\lambda}
    \end{pmatrix}\begin{pmatrix}
        i_1(0)\\
        i_2(0)
    \end{pmatrix}.
\end{equation}
This helps us define the equilibration length $\lambda=1/(2g)$. This is a phenomenological parameter that quantifies mode equilibration. When $x=d\gg \lambda$, the modes are completely equilibrated, which is the limit considered in the main text.


\begin{figure}
    \centering
\includegraphics[width=\columnwidth]{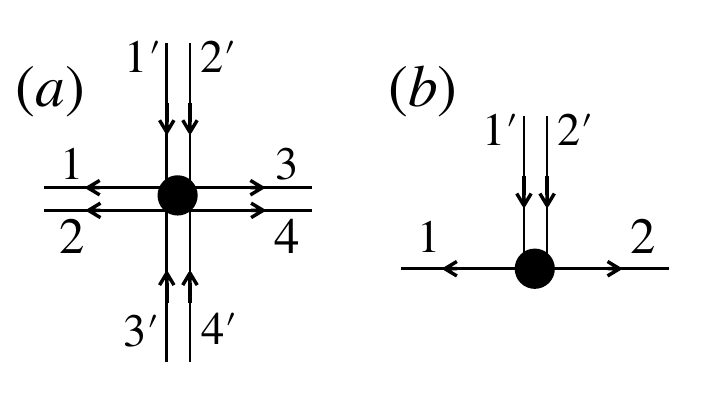}
    \caption{A mode numbering convention for scattering junctions described by scattering matrices in Eqs. \ref{bulk_scatter} and \ref{edge_scatter}. \textbf{(a)} and \textbf{(b)} denote the scattering junctions at the bulk and edge of the square Chern mosaic respectively.}
\label{fig:scattering_junction_numbering}
\end{figure}

Next, we consider mode scattering. For simplicity, we restrict our discussion to the square Chern mosaic, although an extension to a triangular Chern mosaic follows the discussion here straightforwardly. We assign a current to each mode denoted by $i_\alpha(x)$, where $\alpha$ labels the mode, and consider it to be a function of the distance $x$ it has propagated along an edge or domain wall (so that $0\leq x\leq d$)---this implies that the current in each mode can change as it equilibrates with the other mode on that domain wall. Then, generically, at a scattering junction with $m$ modes entering or exiting, we have,
\begin{equation}\label{eq: scattering_equation}
    \begin{pmatrix}
        i_{\alpha_1}(x=0)\\
        \vdots\\
        i_{\alpha_m}(x=0)
    \end{pmatrix}=\mat{S}\begin{pmatrix}
        i_{\alpha_1'}(x=d)\\
        \vdots\\
        i_{\alpha_m'}(x=d)
    \end{pmatrix},
\end{equation}
where $\mat{S}$ is the scattering matrix of the junction. The scattering matrix is real-valued and the sum of the entries on either the rows or the columns is $1$, by the conservation of probability and charge, respectively. The scattering matrix must also be symmetric due to inversion symmetry of the potential. This helps constrain the scattering matrix, so that it is parameterized by only a couple of phenomenological parameters. 

Since the number of modes $m$ at a scattering junction varies based on its position in the sample (at the edge or bulk of the sample), we handle these cases separately. First, we discuss the bulk scattering junctions. We label the modes as in Fig.~\ref{fig:scattering_junction_numbering}(a) and consider two phenomenological parameters---(i) a parameter $t$ that measures the extent of \textit{transmission} between the two modes on the same domain wall \textit{upon} scattering and (ii) a parameter $b$ that measures the \textit{biasing} of the modes towards the left (or right) after scattering. Then, the $4\times 4$ scattering matrix in the bulk of square mosaic is given by the symmetric matrix,
\begin{equation}\label{bulk_scatter}    \mat{S}^{\textrm{bulk}}(t,b)=b\unit_2 \otimes f(t)+(1-b)\mat{X}\otimes f(1-t),
\end{equation}
where $f(t):=t \unit_2 + (1-t)\mat{X}$, $\unit_2$ is the $2\times 2$ unit matrix, $\mat{X}$ is the Pauli-$x$ matrix and $\otimes$ is the Kronecker product. It is easily verified that the sum of entries on the rows and columns is 1. We take $t,b\in [0,1]$ with the limits interpreted as follows---$b=1$ ($b=0$) implies both of the modes completely turn left (right) after scattering while $t=1$ ($t=0$) implies that a mode on a black plaquette scatters completely onto a mode on 
a black (white) plaquette. The limit $b=t=1/2$ is the limit of equal mode scattering in the bulk, which is considered in the main text.


Now, we discuss scattering at junctions on the edge of the sample. Compared to the bulk case, having fewer modes further constrains the scattering matrix. Here, the two modes entering the junction (as labeled in Fig.~\ref{fig:scattering_junction_numbering}(b)) obey a scattering relationship as in Eq.~\ref{eq: scattering_equation} except that the currents on the left hand side are independent of position (since there is a single mode on the edges of the sample). Then, we obtain a $2\times 2$ scattering matrix given by,
\begin{equation}\label{edge_scatter}\mat{S}^{\textrm{edge}}(b)=f(1-b),
\end{equation}
where $b$ has the same definition as above. The matrix is independent of $t$ since only one mode exits on each edge. Other configurations of scattering junctions at the edge can be constructed similarly. We consider the limit of equal branching in the main text here as well, so that $b=1/2$.

The conductance matrix $\mat{G}_{\textrm{eff}}$ for general $(b,t,\lambda)$ can now be obtained from Eqs.~\ref{eq: scattering_equation}, \ref{bulk_scatter}, \ref{edge_scatter} and \ref{mode_equilibration_formula}, since we note that its matrix elements are just the transmission probabilities from lead $i$ to $j$. We leave an explicit discussion of its properties and consequence for transport for future study.

\section{Validity of complete mode equilibration and equal mode scattering}\label{validity}

\begin{figure}
    \centering
    \includegraphics[width=\columnwidth]{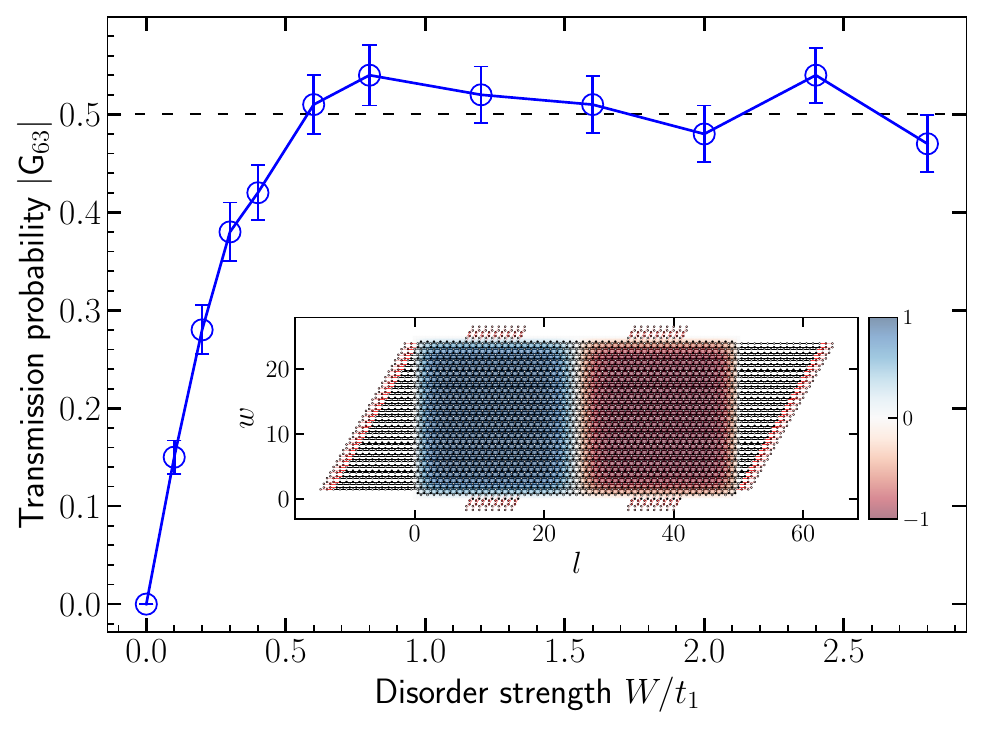}
    \caption{Transmission probability from lead 3 to lead 6 ($=-\mat{G}_{63}$) plotted as a function of disorder strength $W$, obtained from exact transport calculations using \textsc{Kwant}. The Chern mosaic has a single vertical domain wall (see inset and Fig. \ref{fig:examples} E1). The dashed line at $1/2$ indicates the complete mode equilibration and equal mode scattering limit. We choose $t_2/t_1=0.1$ and a Hall bar of size $l\times w = 50\times 25$. Note that for a wide range of disorder---$ 0.6\lesssim W/t_1\lesssim 2.8 $---the transmission probability is $\approx 1/2$. The data is obtained upon averaging over 100 realizations of the disorder potential. \textbf{(Inset)} Kwant realization of the Hall bar of size $l\times w$ in Fig.~\ref{fig:examples} E1. The rectangular region is a honeycomb lattice with the Haldane model Hamiltonian with second nearest neighbor hopping phase $\phi$ (see Eq.~\ref{haldane}), while the leads are decoupled metallic wires. The colorbar indicates the value of $\phi/(\pi/2)$ in real space, as a proxy for the local Chern number.}  
    \label{fig:kwant}
\end{figure}

Our transport computations described in the previous section are agnostic to a microscopic realization of the Chern mosaic phase---all we require are domains with local band structures with non-trivial Chern numbers. We described results in the limit of complete mode equilibration and equal mode scattering. In this section, we start from a microscopic model that may effectively describe a Chern mosaic phase in a two-dimensional material and compute its dc transport numerically. We show that this limit is achieved upon increasing the strength of disorder and persists over a range of disorder strength.

To this end, we construct an effective model of a Chern mosaic using a well-understood microscopic model of a Chern insulator. Specifically, we consider the Haldane model \cite{haldane1988model}, which is given by the following Hamiltonian defined on a honeycomb lattice,

\begin{align}\label{haldane}
\nonumber H=& -t_1\sum_{ \langle ij\rangle }\left[f_i^\dagger f_j+\textrm{h.c.}\right]-\\
&t_2\sum_{\langle ij\rangle\in \alpha, \alpha}\left[e^{-\im\nu_\alpha \phi} f_i^\dagger f_j+\textrm{h.c.}\right]+m\sum_{i\in \alpha, \alpha}\nu_\alpha f^\dagger_i f_i,
\end{align}
where $f_i,f^\dagger_i$ are the usual fermionic annihilation and creation operators at site $i$, $\alpha\in A,B$ is the sub-lattice index and h.c. denotes Hermitian conjugate. $t_1 \: (>0)$ is the strength of the nearest neighbor tunneling, $t_2 \:(>0)$ is the strength of the next nearest neighbor tunneling, $\nu_\alpha\phi$ is a phase accrued in the next-nearest neighbor hopping which is $>0$ for (clockwise) hopping in one sub-lattice but $<0$ for (clockwise) hopping in the other sub-lattice. This cancels the total flux in each hexagonal plaquette. We thus set $\nu_A=1$ and $\nu_B=-1$. This form of next-nearest neighbor hopping breaks time-reversal symmetry, and opens up a topological mass gap at the Dirac points. This competes with the inversion-symmetry breaking term with the staggered onsite energy $m\: (>0)$. Thus, the phases of the Haldane model are given by,
\begin{equation}
 \begin{cases}
&|m| > |3\sqrt{3}t_2\sin\phi|\: : \quad \text{trivial insulator} \\
 & |m| < |3\sqrt{3}t_2\sin\phi| \::  \quad \text{Chern insulator}.
 \end{cases}
\end{equation}
For simplicity, let us consider the $|C|=1$ Chern mosaic with a single vertical domain wall depicted in Fig. \ref{fig:examples} E1. To realize this, it suffices to set $m=0$  and $\phi=\pi/2$ ($\phi=-\pi/2$) to the left (right) of the domain wall\footnote{For the numerical implementation, care is taken so that $\phi$ smoothly interpolates between the domains, yielding a domain wall width $\sim O(\textrm{lattice constant})$, and also a smooth confining potential is applied to the edges of the Hall bar to mask boundary effects.}. 

To study the effects of disorder on the Haldane model Hamiltonian, we also add a disorder potential $H_{W}=\sum_i w_i f_i^\dagger f_i$ where $w_i$ is chosen from a uniform distribution $P(w)= \frac{1}{W}\Theta(\frac{W}{2}-|w|)$. Thus $W$ quantifies the strength of the disorder. Now, we construct a Hall bar using the honeycomb lattice, attach six leads (modeled by decoupled metallic wires) as in Fig. \ref{fig:schematic}(a) (see inset of Fig.~\ref{fig:kwant}) and compute the conductance matrix $\mat{G}$ for transmission between the leads. This is done using the \textsc{Kwant} package \cite{groth2014kwant}, implemented for \textsc{Python}, which employs the wavefunction scattering method to compute the transmission probabilities. 

Disorder only affects the mode equilibration along the domain wall, as well as mode scattering at the junctions on the edge. Thus only the entries $\mat{G}_{53}, \mat{G}_{54}, \mat{G}_{63}$ and $\mat{G}_{64}$ can change as a function of disorder $W$, of which there is only one independent matrix element---say $\mat{G}_{63}$---due to charge conservation and reflection $\times $ time-reversal symmetry about the domain wall. We thus plot the absolute value of $\mat{G}_{63}$ as a function of disorder strength $W/t_1$ in Fig.~\ref{fig:kwant}, averaged over a 100 realizations of the disorder potential. The Hall bar dimensions are $l\times w =50\times 25$ in lattice constant units.  We find that the equal scattering and complete mode equilibration limit is found over a significant range of disorder strength, and achieved at a disorder strength of $W/t_1\approx 0.6$. In the clean limit---at least, in this model and geometry---it appears that a mode originating from the source after scattering into the domain wall gets completely back-scattered to the gray domain towards the source. We note that the Chern insulating phase is intact for the disorder strengths and parameters considered in  Fig.~\ref{fig:kwant} (see for example, Ref.~\cite{gonccalves2018haldane}). Beyond a disorder strength of $\approx 2.8$, we find a breakdown of the network model since all of the matrix elements of the conductance matrix realize non-zero values. 

Upon increasing the width of the Hall bar, we found that $|\mat{G}_{63}|\approx1/2$ is achieved at smaller disorder strengths. This suggests that, at least in this mosaic geometry, disorder primarily controls the equilibration length, and as expected, a larger width also achieves the regime in which the results reported in the main text are valid. 

We also tested the influence of disorder in a mosaic with a single scattering junction in the bulk of the sample. For this, we chose the $m=n=2$ square Chern mosaic (with one vertical and one horizontal domain wall). In this case, we found that the complete mode equilibration and equal mode scattering limit is obtained even in the clean limit ($W=0$), and certainly with disorder. This may be due to increased scattering between modes at the scattering junction; pinpointing the exact cause will require a more detailed study.  

This investigation suggests that for experimental samples with square mosaics and disorder (due to strain, etc.), a semi-classical transport calculation of the Chern mosaic in this simplified limit, as reported, is sufficient.

\section{Higher-order Chern mosaics}\label{app: higher Chern mosaics}

\begin{figure}
    \centering
    \includegraphics[width=\columnwidth]{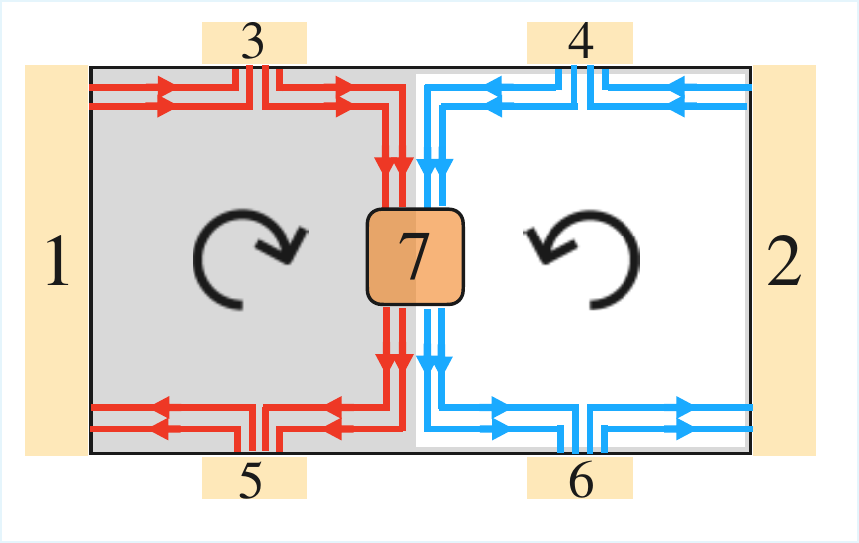}
    \caption{A schematic of a bipartite bipolar higher Chern mosaic with $|C|=2$ and a single vertical domain wall separating the domains. There are 4 chiral gapless edge modes on the domain wall. The physical leads are denoted in light orange, while the auxiliary lead in dark orange. The resistances for this mosaic are provided in Eq.~\ref{higher_chern_mosaic_eq}.}
    \label{fig:higher_chern_mosaic}
\end{figure}

\begin{ruledtabular}
\begin{table*}
 \centering 
\begin{tabular}{cccc}
Number of columns ($n$)& Voltage at lead 4 ($V_4$) & Voltage at lead 6 ($V_6$) & Current ($I$)    \\
\hline
1 & $\frac{1}{3}$ &1 & $\frac{4}{3}$  \\
2  & 0.217391 & 0.478261 & 0.695652 \\
3  & 0.155556 & 0.318519 & 0.474074 \\
4  & 0.119550 & 0.240506 & 0.360056 \\
5  & 0.096683 & 0.193649 & 0.290332 \\
6  & 0.081062 & 0.162183 & 0.243245 \\
7  & 0.069763 & 0.139539 & 0.209303 \\
8  & 0.061224 & 0.122450 & 0.183674 \\
9  & 0.054545 & 0.109091 & 0.163636 \\
10 & 0.049180 & 0.098361 & 0.147541 \\
\end{tabular}
\caption{Voltages at lead 4 $(V_4)$ and lead 6 ($V_6$) and the current $(I)$ for the Chern mosaic with a row of $2n$ full triangular domains ($n$ columns). For $n\geq 2$, the values have been rounded off to six decimal places.}
\label{voltage_currents_small n}
\end{table*}
\end{ruledtabular}

\begin{figure}
    \centering
    \includegraphics[width=\columnwidth]{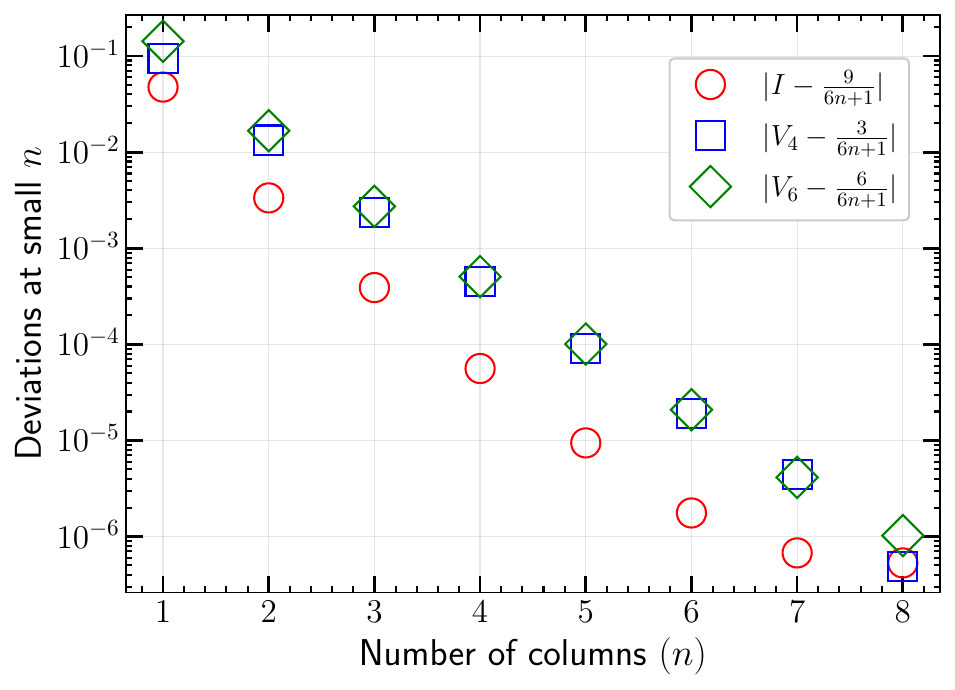}
    \caption{Deviation of the actual voltages (blue square and green diamond) and current (red circle) from the asymptotic value at large $n$ (number of columns) for the Chern mosaic with a row of $2n$ triangular domains (E9).}
    \label{fig:deviationsrow}
\end{figure}

In the main text, we restricted our discussion to Chern mosaics with $|C|=1$. However, our formalism allows for an easy extension to \textit{higher-order} Chern mosaics---Chern mosaics where the Chern number of the domains is $>1$. As an illustrative example, we explain how to use our framework to compute the transport in a Chern mosaic with a single vertical domain wall separating domains with Chern number $|C|=2$. In this case, the domain wall carries 4 chiral co-propagating edge modes, while the physical edges of the Hall bar carry 2 edge modes. We again consider the limit of complete equilibration and equal scattering. Proceeding as in the main text, the effective conductance matrix is now twice the conductance matrix in Eq.~\ref{effective_conductance_matrix}. Using Eq.~\ref{g_formula}, it is easy to see that the reduced conductance matrix is also twice that in Eq.~\ref{reduced_G}. Then, the resistances and the voltage-current vectors are given by,
\begin{align}\label{higher_chern_mosaic_eq}
\begin{split}
&\mathbf{V}=\left(1\: 0\: 1\: 0\: \frac{1}{2}\: \frac{1}{2}\right)^\textrm{T},\;\; I=1,\\
&R_{xx}^{\textrm{top}}=1,\: R_{xx}^{\textrm{bot}}=0,\: R_{xy}^{\textrm{left}}=\frac{1}{2} \;\; \textrm{and}\;\; R_{xy}^{\textrm{right}}=-\frac{1}{2}.
\end{split}
\end{align}
Thus, we find that increasing the Chern number by a factor of 2 decreases the resistances by the same factor---as is expected for a state with a higher Chern number, such as multiple fully filled Landau levels. This behavior is generic for any mosaic geometry, as long as the mosaic is bipolar and bipartite. We can easily generalize---for the Chern mosaic with a single vertical domain wall and domains with Chern number $\pm C$, the resistances and voltage and current vectors are,
\begin{align}
\begin{split}
&\mathbf{V}=\left(1\: 0\: 1\: 0\: \frac{1}{2}\: \frac{1}{2}\right)^\textrm{T},\;\; I=\frac{C}{2},\\
&R_{xx}^{\textrm{top}}=\frac{2}{C},\: R_{xx}^{\textrm{bot}}=0,\: R_{xy}^{\textrm{left}}=\frac{1}{C} \;\; \textrm{and}\;\; R_{xy}^{\textrm{right}}=-\frac{1}{C}.
\end{split}
\end{align}
We expect such higher Chern mosaics may be found in multi-layered 2D material experiments such as in rhombohedral multilayer graphene~\cite{han2024correlated}, where the Chern number of the electronic band scales with the number of stacked layers.

\section{Deviations from large number of column asymptotics for some triangular Chern mosaics}\label{small n deviations}
In this appendix, we provide some details of the finite-size deviations that we obtain for some of the triangular Chern mosaics considered in the main text.  

It suffices to consider resistances of the mosaic with a single row of $2n$ triangular domains (E9 in Table~\ref{main table} and Fig.~\ref{fig:examples}) since the $(m,n)$ triangular mosaics exhibit resistances that may be obtained from the resistances of this mosaic. First, note that the only unknown quantities are the voltages at lead 4 and 6, denoted by $V_4$ and $V_6$, respectively, and the current $I$. $V_3$ and $V_5$ are equal to $1$. In Table~\ref{voltage_currents_small n}, we tabulate these quantities, rounding off to six decimal places. In Fig.~\ref{fig:deviationsrow}, we plot the absolute value of the differences of the voltages and currents with the asymptotic expressions on a logarithmic scale for $n=1,2,\hdots, 8$. Note the rapid convergence onto the asymptotic value so that for $n=8$, the difference is already $O(10^{-6})$. 

\begin{figure}
    \centering
    \includegraphics[width=\columnwidth]{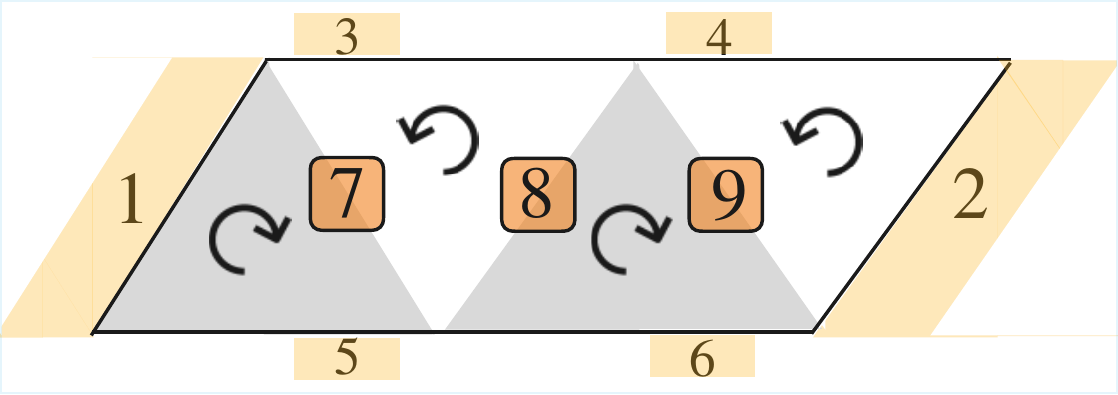}
    \caption{Schematic of a Chern mosaic with $4\times 2$ full triangular domains inside a Hall bar with oblique boundaries, aligned with the triangular axes. The light (dark) orange leads are probe (auxiliary) leads and the round arrows denote the chirality of the edge state in each domain.}
    \label{fig:obliqueHallbar}
\end{figure}

\section{Oblique Hall bar in a triangular Chern mosaic}\label{oblique_hall_bar}
In this appendix, we consider the simplest triangular Chern mosaic geometry with non-zero resistances that are not obtainable as simple fractions. We then rectify this by considering the same geometry but now in a Hall bar whose edges are aligned with the natural axes of the triangular mosaic---an oblique Hall bar in this frame of reference. We find that this Hall bar indeed yields resistances that can be expressed in terms of a simple fraction. 

The geometry we consider is of a single row of $4\times 2$ full triangular domains (or two columns). Note that the geometry with a single row and column $(n=1)$ must have zero longitudinal resistance and so we neglect its discussion. This geometry with the non-oblique (usual) Hall bar is handled by the $n=2$ case of the E9 in Fig.~\ref{fig:examples}, and its voltages and current are provided in Table~\ref{voltage_currents_small n}. Instead, we now consider aligning the left and right boundaries of the Hall bar along the axis of the triangles, as depicted in Fig.~\ref{fig:obliqueHallbar}. Using the rules discussed in the main text, it is straightforward to compute the voltages, current and resistances for this mosaic, and we find,
\begin{align}\label{higher_chern_mosaic_eq}
\begin{split}
&\mathbf{V}=\left(1\: 0\: \frac{4}{9}\: 0\: \frac{2}{3}\: 0\right)^\textrm{T},\;\; I=\frac{8}{9},\\
&R_{xx}^{\textrm{top}}=\frac{1}{2},\: R_{xx}^{\textrm{bot}}=\frac{3}{4},\: R_{xy}^{\textrm{left}}=-\frac{1}{4} \;\; \textrm{and}\;\; R_{xy}^{\textrm{right}}=0.
\end{split}
\end{align}
Nicely, we find that aligning the Hall bar along the triangular axis yields resistances that are expressible as simple fractions unlike in the non-oblique case.



\bibliography{main}




%

\end{document}